# Evaluating electrophysiological and behavioral measures of neural health in cochlear implant users: a computational simulation study

Yixuan Zhang*, Daniel Kipping, and Waldo Nogueira

*Abstract*— *Objective:* **Neural health refers to the condition and functionality of the auditory nerve fibers (ANFs), which are crucial for transmitting sound signals from the cochlea to the brain. However, neural health cannot be directly measured due to current technological limitations. We utilize a computational model to evaluate different indirect methods for estimating ANF neural health.** *Method:* **Two distinct measures for estimating neural health, (i) the threshold levels for focused quadrupolar stimulation and (ii) the change in the electrically evoked compound action potential (eCAP) amplitude growth function for different inter-phase gaps (IPGs), were evaluated in a computational model of an electrically stimulated implanted cochlea. The model combined a 3D finite element method model, an ANF geometry with a realistic spatial distribution, and a neuron model, including an existing phenomenological single-ANF model and an eCAP model. Our experiments simulated different neural health conditions (healthy, shrinked, and degenerated) to model dead regions in different locations of the cochlea.** *Results:* **The results from the computational experiments demonstrated that the threshold levels with focused quadrupolar stimulation were more sensitive to neural health deficits than thresholds with monopolar stimulation. From our data, the difference in threshold levels between quadrupolar and monopolar stimulation seems to be a promising measure of neural health status. However, the results from the eCAP IPG slope and offset effects were not consistently associated with neural health conditions.** *Conclusion:* **Our results suggest that the difference in threshold levels with quadrupolar and monopolar stimulation is a possible method for estimating neural health.** *Significance:* **This study enhances the understanding of neural health and dead regions through a novel computational model, contributing to new approaches for neural health estimation.**

## I. INTRODUCTION

For individuals experiencing severe-to-profound sensorineural hearing loss, cochlear implants (CIs) have achieved immense success by electrically stimulating the surviving auditory nerve fibers (ANFs). Meanwhile, one of the most important challenges researchers face is how the status of ANFs, referred to as neural health, affects the CI's function and the speech perception of CI users. Neural health is compromised by the degeneration of spiral ganglion neurons (SGNs), a phenomenon observed in histological studies of both humans and animals experiencing profound hearing loss [1]. Recent studies also reported that after deafening, not only SGN density, but also the SGN diameter was significantly reduced in humans and animals with severe to profound hearing loss [2] [3]. Estimating the neural health of the auditory nerve, as part of the electrode-nerve-interface (ENI), may help to improve the CI fitting and sound processing and thereby optimize speech perception of CI users. However, due to technical limitations, it is currently not possible to diagnose the neural health status of CI users quantitatively. In recent years, several methods have been proposed to indirectly estimate the neural health status along the cochlea. At least two methods hold potential for neural health estimation: (1) measurement of behavioral thresholds for spatially focused stimulation [4], and (2) evaluation of the effect of inter-phase gaps (IPGs) on the offset or slope of electrically evoked compound action potential (eCAP) amplitude growth function (AGF) [5].

The electrode configuration of CIs influences the spatial voltage distribution in the cochlea and the excitation of the ANFs. As the most commonly used configuration, monopolar (MP) stimulation activates a single intracochlear electrode by sending a current pulse with reference to an extracochlear return electrode [6], [7]. However, stimulation in MP configuration generates a broad voltage spread in the cochlea and activates a large population of ANFs, potentially constraining speech perception. In contrast to MP stimulation, focused stimulation activates multiple intracochlear electrode contacts with different polarities simultaneously to manipulate the spatial spread of voltage to generate a narrower excitation [7-9]. Additionally, in phase stimulation of electrodes has been proposed to steer the current locus of excitation between physical electrode contacts. Different focused stimulation configurations have been proposed in the past decades, including bipolar, partial bipolar, tripolar, quadrupolar (QP), or phase array stimulation configurations [4], [10-12]. By adjusting the configuration coefficients, focused stimulation can produce channels with variable spatial selectivity, which also have different thresholds between channels [13]. Few studies used focused QP stimulation to assess the quality of the

This study is part of the project that received funding from the European Research Council (ERC) under the European Union's Horizon-ERC Program READIHEAR (Grant agreement No. 101044753 – PI: WN) and also funded by the Deutsche Forschungsgemeinschaft (DFG, German Research Foundation) –

SFB/TRR-298-SIIRI – Project-ID 426335750. All authors are with the Hannover Medical School (MHH), Hannover, Germany, and with the Cluster of Excellence Hearing4All, Germany (*Correspondence e-mail: Zhang.Yixuan@mh-hannover.de).

ENI and suggested that focused stimulation is more sensitive to impaired neural health than MP stimulation [13-15].

The eCAP is the synchronized response of the electrically stimulated ANF population and has been considered another measure to estimate neural health status. Several animal studies have demonstrated that animals with higher SGN density tend to have large maximum AGF responses, lower current levels required to evoke an eCAP (defined as the eCAP threshold), and steeper AGF slope [5], [16-18]. Assessed by eCAP measurements on deaf guinea pigs, Prado-Guitierrez et al. (2006) investigated the correlation between eCAP amplitude and SGN density in response to various stimuli with differing pulse durations (PD) and IPG. Specifically, they reported a reduction in eCAP thresholds with a concomitant increase in AGF response amplitudes, resulting in a steeper AGF slope when the input stimuli had a longer IPG. This increase in steepness is referred to as the IPG slope effect, while the shift in eCAP thresholds toward a lower current level is termed the IPG offset effect. However, they also reported a reduction in both effects with lower SGN density, suggesting that both IPG slope and IPG offset effects could be a non-invasive measure of neural health [16]. In another study, Ramekers et al. (2014) measured the IPG threshold and IPG offset effects in guinea pigs with different neural health conditions. Their results also demonstrated that animals with higher SGN density were more sensitive to the IPG change [2]. However, several eCAP studies that evaluated the correlation between neural health and the IPG slope or IPG offset effects led to inconsistent results. Those studies have also shown that the IPG slope and IPG offset effects may be correlated with different aspects of neural health [15, 19, 20-25]. One explanation could be that the scales (linear or logarithmic) of input current or output voltage and the methods used to extract the slope and threshold from the eCAP AGF may impact the estimation of the IPG slope and IPG offset effects. Skidmore et al. (2022) demonstrated that the eCAP AGF slope and threshold are significantly affected by the fitting method used to model the AGF, as well as by the input current unit [26]. Most eCAP experiments commonly adopt two fitting methods: linear and sigmoidal regression. Some computational simulations and animal studies have shown that sigmoidal regression effectively characterizes the AGF within a higher input current range [26-28]. However, most eCAP experiments showed that the AGF at the maximum current tolerable by the subject often does not reach the upper asymptote of the sigmoidal function [29]. Such non-saturating AGFs could cause inaccurate estimation of AGF slope and eCAP threshold. Recently, studies by Brochier et al. suggested the IPG offset effect with logarithmic input current units as a robust neural health measure in humans, while it concluded that the IPG slope effect would be affected by non-neural factors [30], [31]. The results of these experiments may have been influenced by different electrode positions, species, input/output units, fitting methods, and varying degrees of neural health.

In the present study, we developed a high-fidelity computational modeling framework, consisting of a 3D finite element method (FEM) model of an electrically stimulated human cochlea with a CI based on previous work, an ANF model covering the frequency range from 20 Hz to 20000 Hz, a phenomenological single-ANF model and a physiological multi-compartment neuron model [32-34]. With this computational framework, the threshold levels and eCAP responses for different neural health conditions were estimated and evaluated systematically regarding thresholds for MP and focused QP stimulation, as well as the IPG slope and offset effects in eCAP simulations. The present study aims to investigate the influence of neural health conditions in terms of the degree and the size of impaired ("dead") regions. In the following section, we hypothesized that the focused QP stimulation would be more sensitive to impaired neural health conditions than MP stimulation, which would manifest as a higher difference in behavioral threshold levels between the healthy and impaired neural health conditions. In addition, we also hypothesized a correlation between the eCAP IPG slope or IPG offset effect and the neural health conditions, showing more significant IPG slope and IPG offset effects for electrodes surrounded by healthy ANFs when compared to electrodes surrounded by impaired ANFs. Both measures will be statistically assessed regarding their ability to estimate neural health.

## II. Materials and methods

### A. 3D model of electrically stimulated cochlea

#### 1) 3D finite element method model

The FEM model used in this study is an extension of the previously published computational model of the electrically stimulated human cochlea developed in the Auditory Prosthetic Group (see Fig.1a) [34], [35]. For this study, a new CI array (CI632, Slim Modiolar Array, Cochlear Ltd., Sydney, Australia) was inserted into the scala tympani. This CI array was simplified as a spiral cylinder with a linear decreasing radius (0.475 mm at the base and 0.3 mm at the apical end) approximating the dimension and geometry of the CI632 array, consisting of 22 half-band CI electrode contacts distributed over an activate length of 14 mm. According to data from 11 CI632 subjects from Hannover Medical School (MHH), the electrode array was inserted close to the modiolar wall of the scala tympani with an insertion depth of 23.96 mm along the Organ of Corti (OC) re round window (RW) and a distance between the RW and the center of the basal electrode contact of 2.32 mm. Electrode contacts were numbered El to E22 from the basal end to the apical end of the array, and the insertion angle relative to the RW ranged from $25.42°$ to $404.55°$. More details can be found in the Supplementary Material (Suppl. Tab. 1).

As in our previous studies, the conductivity of bone tissue surrounding the cochlea was optimized to match clinically recorded transimpedance values. The transimpedance matrix is a $22*22$ matrix of the intracochlear impedances between each possible pair of stimulating and recording electrodes. Transimpedance matrices were simulated at a range of bone conductivity values and were compared to the average of 11 clinically recorded transimpedance matrices from CI632 subjects. The bone conductivity of the model was optimized to minimize the root mean square error between the simulated and average clinical transimpedance matrices, resulting in a best-fitting bone conductivity of 0.02 S/m.

The electrostatic 3D voltage distribution $V_0^E$ was predicted in an MP configuration: one electrode contact $E$ was activated by delivering a fixed current $I_0 = 1\ \mu A$, while the ground was defined as the entire surrounding bone sphere. Simulations were performed with COMSOL Multiphysics v5.6 (COMSOL AB,

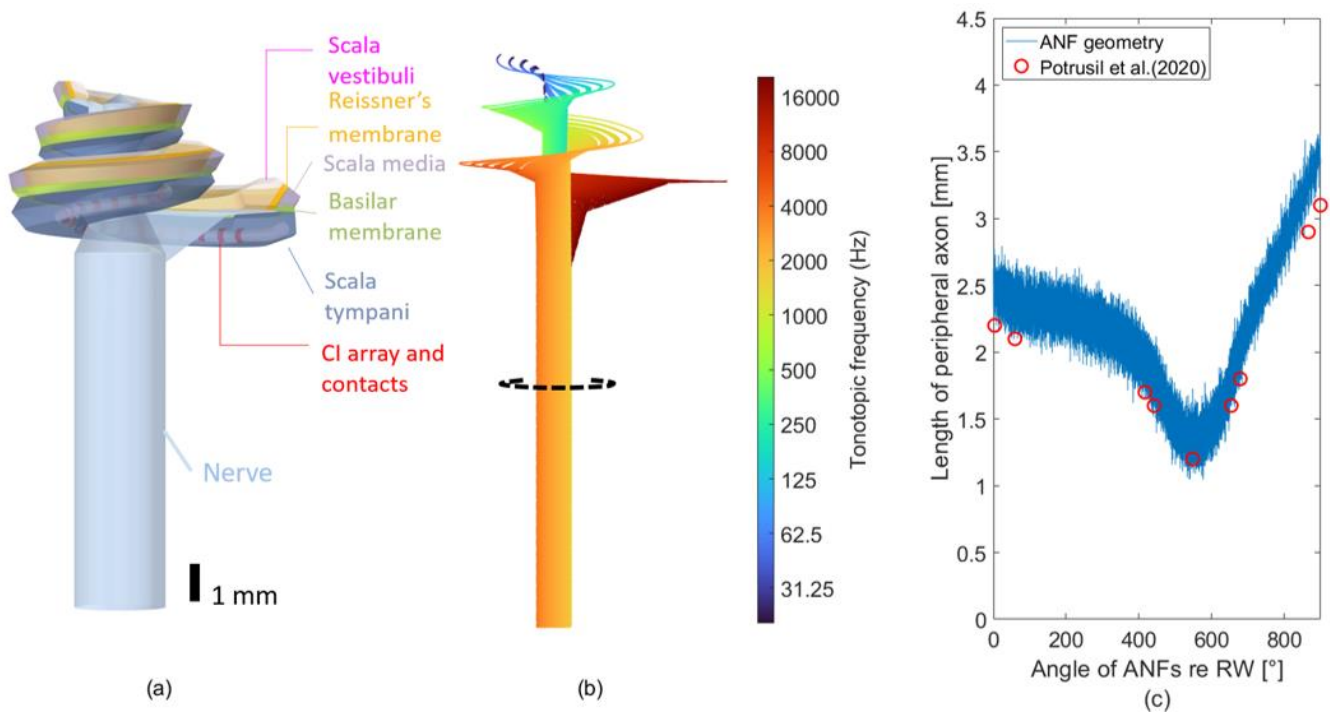

Fig. 1. 3D reconstruction of a human cochlea and auditory nerve fiber (ANF) geometry. (a) The 3D human cochlea geometry contained a CI electrode array in the scala tympani and was embedded in a bony sphere with a diameter of 35 mm. (b) 30000 ANFs were modeled inside the nerve geometry (black dashed circle) covering the frequency range from 20 Hz to 20 kHz to sample the voltage distribution. (c) Length of the peripheral axons against ANF angle re. round window (RW) compared with data of Potrusil et al (2020) [36].

Stockholm, Sweden) in an Intel Core i9-9900k workstation with 32 GB of RAM, and the simulation process was controlled from MATLAB 2023a (The MathWorks, Inc., Natick, USA) using a custom script through the LiveLink interface between MATLAB and COMSOL.

### 2) 3D ANF geometry and neural health conditions

This study incorporated updated realistic ANF trajectories through the modiolus and Rossenthal's canal updating the proposed FEM and ANF model used in [34]. The range of characteristic frequencies was updated and ranged from 20 Hz to 20 kHz to match the normal human frequency range. According to the study by Potrusil et al. (2020), the somata placement and the length of the peripheral process of each ANF vary along OC. Their data were used to form the position of the soma for each ANF, as shown in Fig.1c. As in [34], the bipolar ANFs were subdivided into compartments representing the peripheral terminal, the nodes of Ranvier, a pre-somatic region, and the soma. The internodal regions between the nodes of Ranvier were shielded by fully insulated myelinated segments. The segment-mapping process for each ANF was identical to [34], using the ANF morphology proposed by Kalkman et al. [37] (see Fig.2a).

Each ANF was simulated in one of the following three states (see Fig. 2b):

- 'Healthy': Fully healthy ANF.
- 'Shrinked': ANF with 50% reduced diameter in the peripheral axon.
- 'Degenerated': ANF with entire loss of the peripheral axon.

The states were assigned to the ANFs based on the overall neural health condition as follows. The healthy neural condition was defined as 'healthy' for all ANFs. In addition, conditions with impaired neural health were defined that contained a dead region, i.e. a region with impaired ANFs, at a certain place in the cochlea. The 12 dead regions had an impaired length (IL) of either 0.5 mm, 1.5 mm, or 5.0 mm, and were spaced at intervals of 30 degrees along the OC. In each condition with impaired neural health, either 'Shrinked' or 'Degenerated' ANFs were assigned to one of these regions (see Fig.2c). All simulated neural health conditions are listed in Supplementary Material (Suppl. Tab. 2).

The ANF model geometry was used to spatially sample the 3D voltage distribution at the compartments $i$ of the ANFs $j$, resulting in voltage matrices $V_{0,i,j}^E$. We assumed that capacitive effects of tissues and electrodes could be ignored in the FEM model. Therefore, the time-dependent voltage distribution $V_{i,j}^E(t)$ for an arbitrary time-varying stimulus current $I(t)$ could be predicted by scaling the quasi-static solution:

$$V_{i,j}^E(t) = I(t) \cdot V_{0,i,j}^E / I_0. \tag{1}$$

### B. Experiment 1: Behavioral thresholds of QP vs. MP stimulation

#### 1) Neural excitation for QP and MP stimulation

A phenomenological single-ANF model based on the adaptive integrate-and-fire point neuron model of Joshi et al. (2017), with the modifications introduced by Kipping et al. (2022), was implemented to simulate direct electrical stimulation and predict the spike generation of each ANF [33], [38]. As our study focused on electric-only stimulation, we utilized the uncoupled version of Kipping's electric-acoustic model, implementing electric input stimuli $I_j(t)$ for ANF $j$ and setting acoustic input and output to zero.

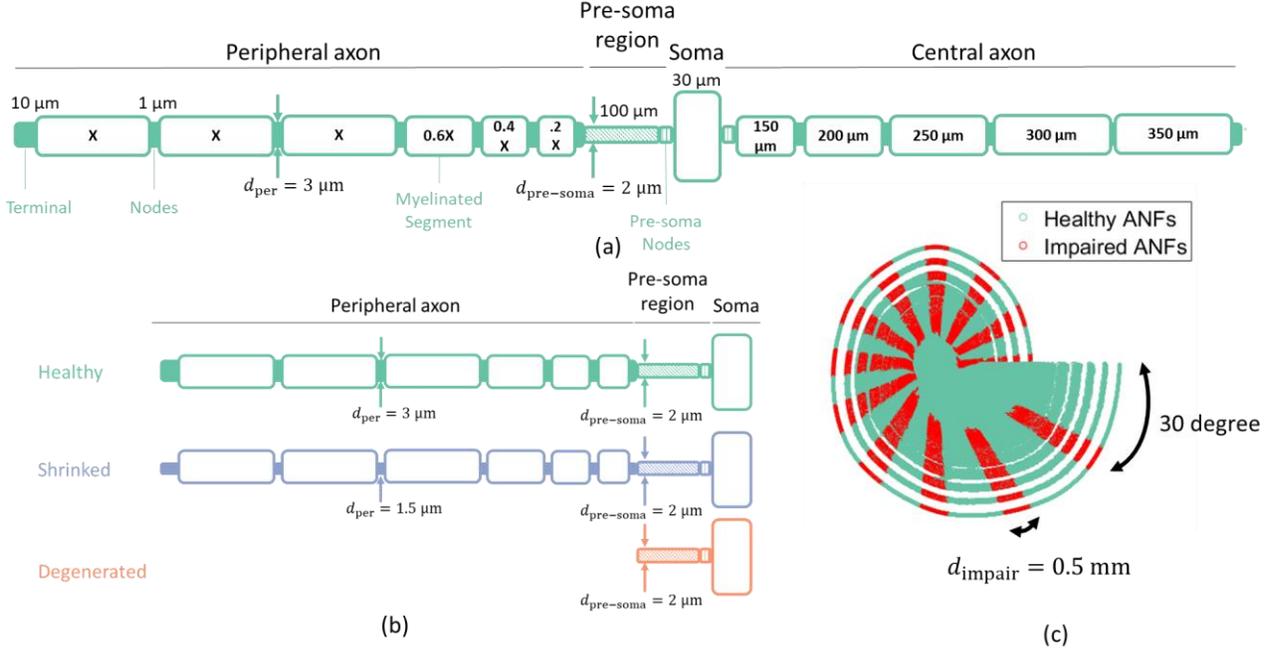

Fig. 2. Auditory nerve fiber (ANF) morphology and different neural health conditions. (a) The ANF morphology was based on Kalkman et al. (2014) [37]. Each ANF was segmented into the peripheral axon, the pre-soma region (box with diagonal stripes), the pre- and post-soma nodes (boxes with vertical stripes), the soma, and the central axon. The base length x and the number of myelinated segments in the peripheral axon were modified based on the length of the peripheral axon. (b) Schematic representation of an ANF in different neural health conditions (Healthy, Shrinked, and Degenerated). The ANF impairment affected only the peripheral axons, whereas the pre-somatic and somatic regions as well as the central axons remained in the original state. (c) Regions of local ANF impairment, spaced at intervals of 30 degrees. The impaired regions covered a length of either 0.5 mm, 1.5 mm, or 5.0 mm along the Organ of Corti. For each simulation, the ANFs were impaired ('shrinked' or 'degenerated') in one of the 12 regions.

As in [34], the current study used the scaling factor $s_j^E$ to build the spatial connection between the 3D voltage distribution $V_{0,i,j}^E$, and the ANF phenomenological model. For simplicity, the ANF geometry selects only one node to represent activity in each ANF caused by an input stimulus. Following the idea of [39], this selected node is the most likely to generate a spike in response to an input stimulus, assuming that this is the node that has the largest induced current by the extracellular electric voltage $V_{0,i,j}^E$. The induced current of each compartment $i$ in an ANF $j$ is estimated with the activating function[40]:

$$\text{AF}_{i,j}^E = G_{i-1,j} \cdot (V_{0,i-1,j}^E - V_{0,i,j}^E)/C_{i,j} - G_{i,j} \cdot (V_{0,i,j}^E - V_{0,i+1,j}^E)/C_{i,j}, \quad (2)$$

where $C_{i,j}$ is the membrane capacitance and $G_{i,j}$ is the axonal conductance. The selected node $i_{\max}^E(j)$ has the largest absolute activating function values along the fiber:

$$i_{\max}^E(j) = \text{argmax}_i(|\text{AF}_{i,j}^E|). \quad (3)$$

Equation (3) excludes nodes in pre-somatic and somatic regions, as well as the nodes adjacent to the somatic region. This exclusion is necessary because these nodes exhibit abnormally high $\text{AF}_{i,j}^E$ with the inconsistent $G_{i,j}$ value.

In addition, a normalization factor $\lambda$ was introduced to adjust the scaling factor values for all electrodes, which ensures the preservation of the scaling factor as average input and resulted in averaged excitation profile across all electrodes in the model:

$$\lambda = \frac{1}{N_E} \cdot \frac{1}{N_{ANF}} \cdot \sum_E \sum_j \left| \frac{1}{\text{AF}_{i_{\max}^E(j),j}^E} \right|, \quad (4)$$

where $N_E = 22$ is the total number of electrodes and $N_{ANF}$ represents the total number of ANFs. Notice that the normalization of Equation (4) was based on the $\text{AF}_{i,j}^E$ in Healthy conditions with MP stimulation mode. The same normalization factor was applied to the scaling factors in all neural health conditions for both MP and QP stimulations.

The scaling factor $s_j^E$ and the input stimuli $I_j(t)$ for each fiber were computed as:

$$s_j^E = \lambda \cdot \text{AF}_{i_{\max}^E(j),j}^E, \quad (5)$$
$$I_j(t) = I(t) \cdot s_j^E. \quad (6)$$

The single-fiber threshold $I_{\text{thr},j}^E$ of ANF $j$ for MP stimulation of electrode $E$ was defined as the stimulation level that elicited a firing efficiency of 50% in this ANF (i.e. ANF $j$ had a probability of 50% to respond with a spike to a single pulse presented at the level $I_{\text{thr},j}$). Without the ENI (i.e. $s_j^E = 1$ in Equation (6)), the point-neuron model had a certain threshold value $I_{\text{thr},0}$ that depended on the pulse shape (e.g., phase duration and IPG) and was determined iteratively. Following Equation (6), the single-fiber thresholds for all ANFs could be derived from this value as:

$$I_{\text{thr},j}^E = I_{\text{thr},0}/\text{abs}(s_j^E). \quad (7)$$

These single-fiber threshold currents were used to predict the T-levels and M-levels, which will be described in detail in the following section.

The focused QP stimulation followed the idea of Bierer et al. (2015), stimulating four adjacent electrodes simultaneously controlled by two coefficients: the steering coefficient $\alpha$ and the focusing coefficient $\sigma$. A combination of four electrodes with specific coefficients $\alpha$ and $\sigma$ defined a QP channel. The two center electrodes acted as active electrodes, and the two flanking (outer) electrodes served as the return electrodes. The coefficient $\alpha \in [0,1]$ steered the current between the two center electrodes, sending current partially through one active electrode (amplitude $\alpha$) and the remaining current to the other one (amplitude $1 - \alpha$). The focusing coefficient $\sigma$ specified the return current through the two flanking electrodes (amplitudes $-\sigma/2$) [4].

Due to the linear relation of the FEM simulation with respect to the injected current, the voltage distribution of a QP channel $V_{QP,0,i,j}^{E}$ can be obtained as a linear combination of the MP voltage distributions $V_{0,i,j}^{E}$ of the four adjacent electrodes that form the channel:

$$V_{QP,0,i,j}^{E} = -\frac{\sigma}{2} \cdot V_{0,i,j}^{E-1} + \alpha \cdot V_{0,i,j}^{E} + (1 - \alpha) \cdot V_{0,i,j}^{E+1} \\ -\frac{\sigma}{2} \cdot V_{0,i,j}^{E+2}, \quad (8)$$

where $E \in [2,20]$ is the basal center electrode of the QP channel (i.e., the center electrode located more basal than the other one). The channel number for QP channels was defined as "$E + \alpha$", roughly representing the location of the peak voltage relative to the physical electrode contacts. For example, channel 2.5, with the basal center electrode $E = 2$ and $\alpha = 0.5$, involved stimulation of the electrode contacts 1 through 4, with the current equally distributed between the two central electrodes E2 and E3. The special cases $\alpha = 1.0$ and $\alpha = 0.0$ were labeled with an additional suffix 'a' or 'b'. For instance, both channels 3.0a (with $\alpha = 1.0$ and $E = 2$) and 3.0b (with $\alpha = 0.0$ and $E = 3$) directed all current through electrode E3. The steering coefficient $\alpha$ ranged from 0 to 1 in 0.1 steps, while the focusing coefficient $\sigma = 0.9$ was fixed, following the same configuration as in [4]. The scaling factors resulting from QP stimulation were estimated using Equations (5) and (2) where the MP voltage distribution $V_{i,j}^{E}$ was replaced by the QP voltage distribution $V_{QP,i,j}^{E}$ from Equation (8).

### 2) Estimation of T-levels and M-levels

The behavioral threshold (T-level) and the behavioral most comfortable loudness level (M-level) for both MP and QP stimulation were estimated as the current levels that caused a 1 mm or a 4 mm spread of excitation along the OC, respectively [41], [42]. For this purpose, each ANF was considered as excited when the stimulating current was larger than its single-fiber threshold $I_{thr,j}^{E}$ in Equation (7). T- and M-level differences were calculated between Healthy and impaired conditions by subtracting the corresponding current levels. Some channels exhibited increased current levels under varying neural health conditions (e.g., elevated T-levels and M-levels, or inconsistent AGFs) due to their proximity to regions of nerve damage. These channels were classified as affected channels. The maximum increases in T-levels between impaired and Healthy conditions ($\Delta T_{NH}$) across all affected channels for both stimulation modes were used to evaluate the sensitivity of T-levels on the neural health condition.

The input stimulus for focused threshold experiments was a train of cathodic-leading pulses with a phase duration of 98 μs and an inter-phase gap of 4 μs, presented at 997.9 pulses per second. The pulse train was configured with a duration of 200.4 ms and a subsequent silence period of 300 ms.

### C. Experiment 2: ECAP IPG slope and offset effects

#### 1) eCAP simulation

This experiment was based on the computational modeling framework for CAPs proposed by [33]. This framework assumes that the eCAP response elicited by electrode $E$ and recorded at electrode $R$ can be predicted by individual single-fiber CAP contributions (SFCCs) and a peri-stimulus time histogram (PSTH). To quantify the contribution of a single ANF to the measured CAP, the SFCC was simulated with a nonlinear multi-compartment neuron model coupled to the 3D FEM model. This coupled neuron-FEM model was used to simulate the generation and propagation of action potentials along the ANF, the resulting transmembrane currents of the ANF, and the measured potential in the recording electrode $R$. Meanwhile, the aforementioned phenomenological integrate-and-fire model (see Section II.B.1) predicted the resulting PSTH$_j$ with the electrical pulse input, which represented the spike times of ANF $j$. A detailed description of the eCAP simulation can be found in [34]. With this assumption, the eCAP response is estimated as

$$eCAP(t) = \sum_j \left( PSTH_j * SFCC_j^{(E,R)} \right)(t), \quad (9)$$

where the sum adds the contributions from all ANFs $j$.

ECAPs were predicted for MP mode and biphasic cathodic-leading single pulses with two different IPGs (short IPG: 8 μs; long IPG: 40 μs). The phase duration of both pulses was 42 μs. To allow for time-efficient eCAP simulations, the number of ANFs was down-sampled from 30000 to 3000 for experiment 2. The down-sampling was compensated by multiplying all eCAP responses by a factor of 10, as proposed by [34]. The amplitude of the eCAP was measured as the difference between the negative peak N1 and the positive peak P1 of the eCAP waveform at each stimulated level.

#### 2) eCAP amplitude growth function measurement

Before predicting the eCAP AGF, the model framework estimated the T-levels and M-levels with different neural health conditions and stimuli (see Section II.C.1). The stimulation current range to estimate the eCAP AGF started from 5 dB below the T-level and ranged up to M-level in steps of 1 dB for each stimulating electrode $E$. The recording electrode $R$ was defined as the stimulating electrode $E$ plus two (+2) in the apical direction. Limited by the total number of electrodes, the stimulating electrode ranged from 1 to 20. To avoid the influence of non-saturating AGFs below the M-level, the AGF fitting process was characterized by a linear regression implemented with logarithmic input units (dB re 1 μA) and linear output units (μV) [23], [31], [43]. The intercept of this linear regression line with the input current axis was defined as the eCAP threshold, while its slope was defined as the eCAP AGF slope. AGF slope and eCAP threshold were contrasted and analyzed between two IPGs to assess the IPG slope and IPG offset effects. Note that only the input current levels eliciting eCAP amplitudes larger than zero were utilized for the linear regression fitting and estimation of AGF slopes and thresholds. Furthermore, only channels that showed a change in the AGFs between impaired and Healthy conditions were considered as affected channels for the analysis of the influence of neural health conditions on the IPG slope and IPG offset effects. Due to the generally higher sensitivity of the linear slope to data variation, a change in the AGF was characterized by a distinct simulated eCAP threshold with at least a 1% deviation from the Healthy condition or an AGF slope with at least a 5% difference from the Healthy condition. Differences in AGF slope and eCAP threshold of one stimulation-recording electrode pair were calculated as the variation between the short IPG (8 µs) and the long IPG (40 µs). For each specific combination of poor neural health (i.e., neural health condition and IL), the minimum difference in AGF slope and eCAP threshold across affected channels for each neural health condition and IL were marked as ΔSlope and ΔThreshold, respectively. These differences were then compared to the corresponding ΔSlope and ΔThreshold values recorded for the same affected channels under Healthy conditions as the reference.

### D. Statistical analysis

All data were processed and analyzed via MATLAB R2023a. In experiment 1, ($\Delta T_{NH}$ for 12 locally impaired regions and 3 ILs were evaluated using an independent student t-test. Simulated data exhibiting no disparity between impaired and Healthy conditions (i.e., $\Delta T_{NH} = 0$) were excluded from the analysis. A Bonferroni adjustment for four comparisons ($\Delta T_{NH}$ in two distinct impaired conditions and two stimulation modes: MP Shrinked−Healthy, MP Degenerated−Healthy, QP Shrinked−Healthy, and QP Degenerated−Healthy) was applied, resulting in an adjusted significance level of 0.0125.

## III. RESULTS

### A. Experiment 1: Behavioral thresholds of QP vs. MP stimulation

Fig. 3 illustrates the T-levels for MP and QP stimulation when ANFs were impaired in Region 6 (approximately 170°–190° relative to the RW), with an IL of 0.5 mm. Estimated behavioral thresholds in the other impaired regions and lengths can be found in the Supplementary Material (Suppl. Fig.1–3). The T-levels for QP stimulation across all channels showed a similar pattern as the T-levels for MP stimulation. However, for healthy ANFs the T-levels for MP stimulation were on average 7.35 dB lower across all channels, consistent with previous studies [4], [14]. In general, impairing the ANFs in a local dead region increased the T-levels in the vicinity of this region. This increase became more pronounced when the size of the impaired region increased from 0.5 mm to 1.5 mm and 5.0 mm (Suppl. Fig. 1(a-c)). No discernible difference in the T-level increase for both modes was observed between the impaired condition groups Shrinked and Degenerated with an IL of 0.5 mm, on average 1.13 dB across all affected channels. However, a distinct difference emerged for QP T-levels when the IL was increased to 1.5 mm. Compared to the MP mode, focused QP T-levels show a higher and more discernible increase for different neural health conditions. For the Shrinked condition, the average T-level increases relative to the Healthy condition were 2.02 dB at 0.5 mm IL, 3.29 dB at 1.5 mm IL, and 4.92 dB at 5 mm IL. For the Degenerated condition, the average T-level increases were 2.30 dB at 0.5 mm IL, 5.19 dB at 1.5 mm IL, and 10.38 dB at 5.0 mm IL.

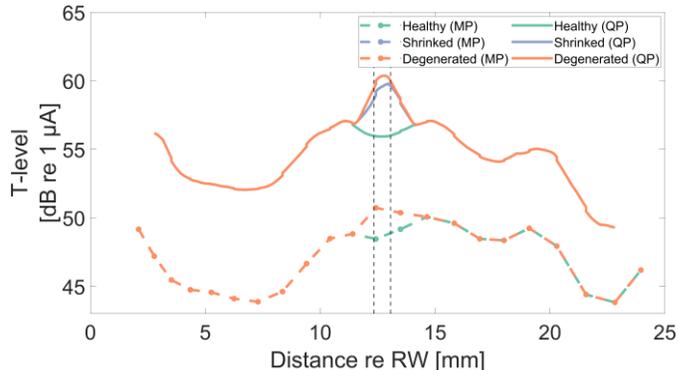

Fig. 3. Example of estimated behavioral T-level profiles for monopolar (MP) and quadrupolar (QP) stimulation in different neural health conditions applied to region 6 over an impaired length of 0.5 mm. The x-axis shows the angle of insertion of each MP or QP channel relative to the round window (RW). The dotted black vertical lines indicate the borders of the impaired region. The QP stimulation results are displayed as a moving mean value average across 11 neighboring channels.

Fig.4(a-c) shows the maximum increase in T-levels across all channels ($\Delta T_{NH}$) from the Healthy to an impaired neural health condition for MP and QP stimulation and three ILs. For a relatively small IL of 0.5 mm, the $\Delta T_{NH}$ with QP stimulation was significantly larger than the $\Delta T_{NH}$ with MP stimulation, both for the Shrinked and the Degenerated condition ($p <$ 0.00025; see Fig.4a). However, there was no significant difference between the Shrinked and the Degenerated conditions when using the same stimulation mode (MP, $p = 1$; QP, $p = 0.0423$). As the IL increased in the Shrinked condition, the difference in $\Delta T$ between the two stimulation modes decreased. The $\Delta T_{NH}$ for both modes approached the same value of 6.02 dB at the largest IL of 5.0 mm (see Fig.4c). Meanwhile, for ILs 1.5 mm and 5.0 mm, significant differences in $\Delta T_{NH}$ between the Shrinked and Degenerated conditions were observed for QP stimulation ($p < 0.00025$). In contrast, MP stimulation showed a significant difference between the Shrinked and the Degenerated conditions only at the IL of 5.0 mm (IL= 1.5 mm, $p = 0.0349$; IL= 5.0 mm, $p = 0.0013$).

From the T-level profiles (Fig. 3 and Suppl. Fig. 1-3) and maximum T-level elevations (Fig. 4a-c) it is apparent that the T-levels for QP stimulation responded with larger elevations to local ANF impairment than the T-levels for MP stimulation. Based on this observation, we hypothesized that the difference in T-levels between the QP and MP stimulation modes could be used as a measure of the neural health state, with larger QP-MP differences indicating a more severe ANF impairment. Therefore, an additional test was performed comparing the maximum T-level differences between the QP and MP stimulation modes across all channels within the same neural health condition. Given the natural fluctuations in T-levels across channels, which were attributed to variations in the 3D model (e.g., insertion position of the CI array, differing widths and heights of the scala tympani in basal versus apical regions), the difference in QP-MP T-levels within healthy condition exhibited a decline from the base to the apex. To correct this decline, a linear regression model was fitted to the T-level differences between the QP and MP stimulation modes along the insertion distance relative to the RW of each channel. For this purpose, the data points from the 22 MP channels were up-sampled by linear interpolation to match the number of data points for the QP channels. The residuals from this model were analyzed to account for these fluctuations. Specifically, the maximum residuals of each impaired condition, defined as $\Delta T_{mode}$, are shown as boxplots in Fig.4(d-f). For reference, the maximum T-level difference between the QP and MP stimulation modes in the Healthy condition, determined as $\Delta T_{mode} = 1.53$ dB from the residuals of the corresponding regression model, is indicated as the horizontal line.

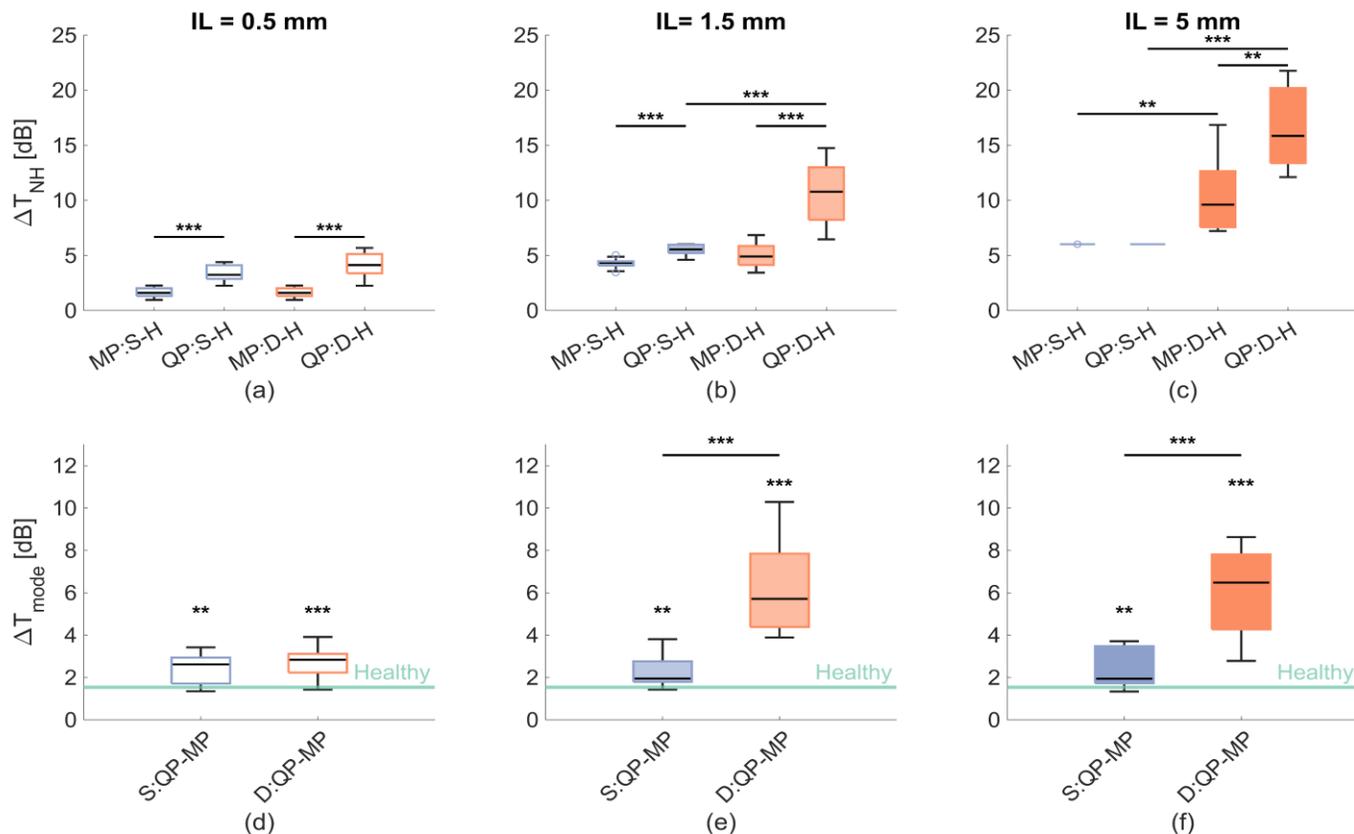

Fig. 4. (a-c) Maximum increase in T-level ($\Delta T_{NH}$) across all channels between an impaired neural health condition (Shrinked "S" or Degenerated "D") and the Healthy case ("H") when the neural health impairment was applied over an impaired length (IL) of 0.5 mm (a), 1.5 mm (b), or 5.0 mm (c). The stimulation mode was either monopolar ("MP") or quadrupolar ("QP"). (d-f) Maximum residual difference in T-level ($\Delta T_{mode}$) between the QP and MP stimulation modes as determined from a linear regression model fitted to the QP−MP differences (see main text for details), for ILs of 0.5 mm (d), 1.5 mm (e), or 5.0 mm (f). The value for the healthy case is indicated by the horizontal line, and values for the impaired neural health conditions (Shrinked "S" or Degenerated "D") are represented by the boxplots. Color coding: Orange represents degenerated, blue represents shrinked. (Significance levels for (a-c): *** for p < 0.00025, ** for p < 0.0025 and * for p < 0.0125; Significance levels for (d-f): *** for p < 0.001, ** for p < 0.01).

To statistically assess the $\Delta T_{mode}$ difference between the Healthy and impaired conditions, a one-sample t-test was applied to the maximum residuals for each IL. Significant $\Delta T_{mode}$ differences between impaired and Healthy conditions were observed for all impaired conditions and IL (Shrinked: QP-MP, $p <0.01$; Degenerated: QP-MP, $p <0.001$). Simulation results also showed a significant difference between Shrinked and Degenerated conditions with IL =1.5 mm and 5.0 mm (both IL= 1.5 mm, and IL= 5.0 mm, $p < 0.001$). However, the channels represented by $\Delta T_{mode}$ for impaired conditions were not always situated within the impaired region (see Suppl. Fig.4). When ANFs were in poor neural health conditions with small ILs or in Shrinked condition, $\Delta T_{mode}$ may refer to channels far from the impaired region.

### B. Experiment 2: ECAP IPG slope and offset effects

Table 1 presents the average T-levels and M-levels across all stimulated electrodes (E1-E20) and across 12 impaired regions, estimated using single pulse stimuli with an IPG of 8 μs (IPG-8) and 40 μs (IPG-40). Similar to other studies, the estimated T-levels and M-levels through the phenomenological model were lower for the long IPG stimuli than for the short IPG [16], [17], [21]. Nevertheless, both average T-levels and M-levels were relatively homogenous across all neural health conditions, with variations of less than 0.5 dB for ILs of 0.5 or 1.5 mm. Fig. 5 shows an example of how eCAP AGFs were simulated for different IPGs and neural health conditions. This example involves stimulation on electrode E1, recording on electrode E3, with neural health impairment applied to region 1 over an IL of 5.0 mm. Fig. 5(a-d) show the compound PSTH of all ANFs. The spike counts changed depending on the distinct IPG and neural health condition, along with the corresponding T-levels and M-levels. Fig. 5e illustrates the SFCCs generated by an ANF (situated 30° re. RW) under varying neural health conditions and IPGs. The predicted eCAP waveforms shown in Fig.5(f-g) had a biphasic morphology similar to the clinically recorded response, consisting of a negative N1 peak followed by a positive P1 peak within a time window of less than 0.5 ms. The predicted eCAP amplitude increased with increasing stimulation level as shown in Fig. 5h on a logarithmic-linear scale. Further AGFs resulting from the different neural health conditions, IPGs, and stimulating and recording electrodes can be found in the Supplementary Material (Suppl. Fig. 3).

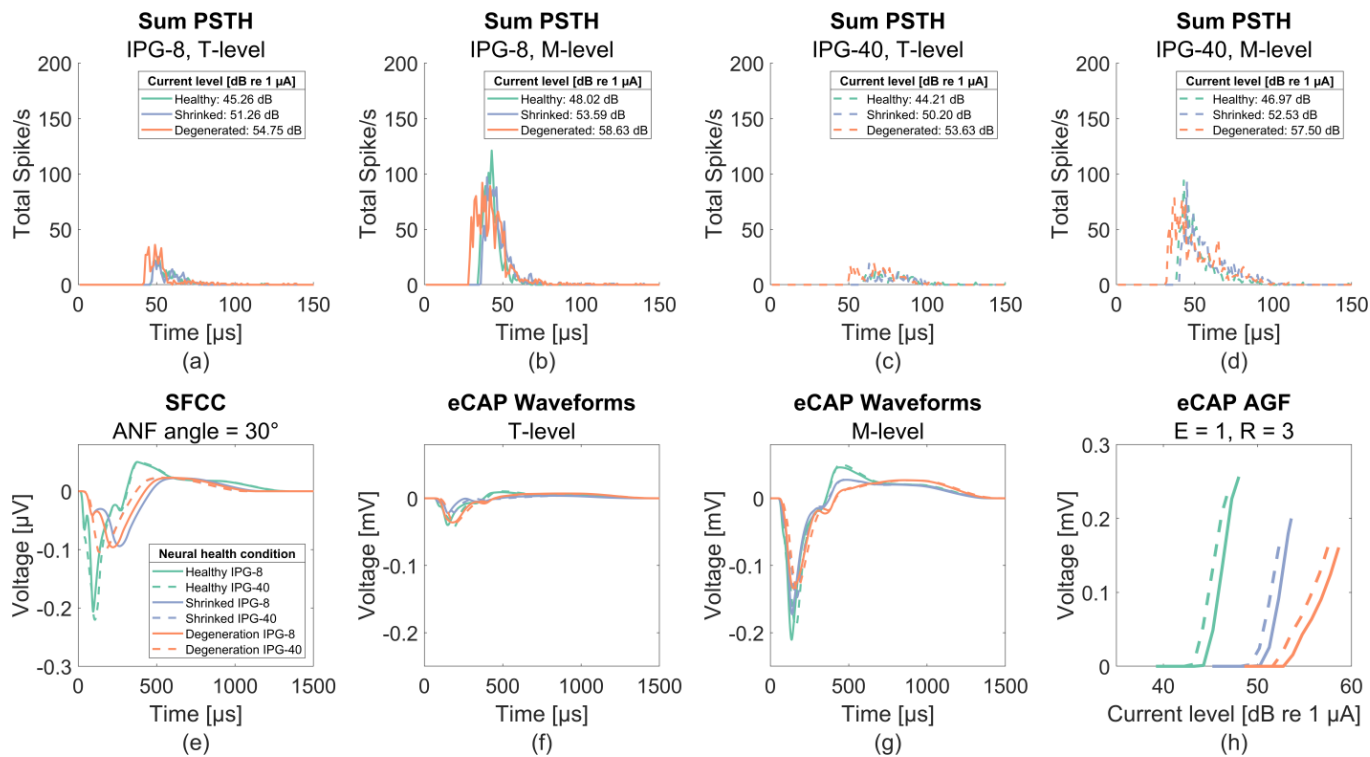

Fig.5. Examples of the peri-stimulus time histograms (PSTHs), single-fiber CAP contributions (SFCCs), electrically evoked compound action potentials (eCAPs), and eCAP amplitude growth function (AGFs) for different neural health conditions and inter-phase gaps (IPGs). In this example, the stimulating electrode E1 and the recording electrode E3 were positioned at 24.20° and 40.61° relative to the round window (RW), respectively, and the neural health impairment was applied to region 1 over an impaired length of 5.0 mm (0.37°-72.4 ° re. RW). (a-d) Compound PSTHs of 3000 auditory nerve fibers (ANFs) in response to single pulse stimuli with IPGs of 8 μs (IPG-8) or 40 μs (IPG-40) presented at the corresponding T-level or M-level for each distinct neural health condition. (e) Estimated single fiber CAP contributions (SFCCs) in response to the same stimuli for an ANF positioned at approximately 30° relative to the RW. (f-g) eCAP waveforms estimated at T-level and M-level for the different neural health conditions. The legend is identical to subplot (e). (h) Predicted eCAP AGFs ranging from T-level to M-level for the different neural health conditions and IPGs. The legend is identical to subplot (e).

As mentioned before, the eCAP threshold and eCAP AGF slope were fitted using linear regression on the logarithmic-linear AGF from the current level generated response (i.e., the amplitude of eCAP>0) to the M-level of the corresponding electrode. Data points that did not yield an eCAP amplitude were excluded from the regression to ensure the accuracy of the linear fit. These two characteristics were contrasted and analyzed between the short and the long IPG to assess the IPG slope (ΔSlope) and IPG offset (ΔThreshold) effects [5], [17]. Fig. 6(a-d) presents the linear fitting of AGFs corresponding to stimulated electrodes E8, E10, E12, and E14, selected as part of the affected channels across three neural health conditions, with an IL of 5.0 mm in Region 6. It was observed that poor neural health condition, compared to Healthy condition, had two distinct effects on the corresponding AGF morphology of affected channels: (1) it weakened the maximum AGF amplitude without notably affecting the eCAP threshold (as shown in Fig.6a); (2) it impacted both AGF slope and eCAP threshold (as shown in Fig.6(b-d)). Meanwhile, with increasing IPG, an increase in the eCAP threshold of approximately 1 dB was observed across most channels, regardless of nerve health conditions. The across-channel differences in AGF linear slope and eCAP threshold between two IPGs for this specific neural health condition are shown in Fig.6(e-f). The complete results for across-channel differences in AGF linear slope and eCAP threshold between the two IPGs under different neural health conditions are provided in the Supplementary Material (Suppl. Fig.4-5). The AGF slope and eCAP threshold differences across channels were highly variable, and simulated results failed to reveal a clear pattern relating ΔSlope or ΔThreshold to neural health conditions for each stimulating electrode E.

## IV. DISCUSSION

This study investigated potential measures of neural health (MP vs. QP T-levels, and eCAP IPG slope and IPG offset effects) within an updated 3D computational model of an electrically stimulated cochlea. This model was coupled with a physiological multi-compartment neuron model and a phenomenological integrate-and-fire neuron model to predict the individual neural response of ANFs with different neural health conditions. The measures were evaluated in terms of their ability to detect "dead regions", i.e., local cochlear regions with poor neural health. Dead regions were simulated with different sizes (ILs of 0.5 mm, 1.5 mm, and 5.0 mm) and severities ("Shrinked" ANFs with reduced diameter and "Degenerated" ANFs with complete loss of the peripheral axon).

TABLE I
AVERAGE BEHAVIORAL T-LEVELS AND M-LEVELS FOR EXPERIMENT 2 [dB re 1μA]

| NEURAL HEALTH CONDITION | IPG-8 | | IPG-40 | |
|---|---|---|---|---|
| | T-level | M-level | T-level | M-level |
| Healthy | 43.32 | 50.08 | 42.27 | 49.02 |
| Shrinked (0.5 mm) | 43.43 | 50.18 | 42.38 | 49.12 |
| Shrinked (1.5 mm) | 43.82 | 50.35 | 42.76 | 49.29 |
| Shrinked (5.0 mm) | 45.24 | 50.94 | 44.17 | 49.88 |
| Degenerated (0.5 mm) | 43.43 | 50.24 | 42.38 | 49.12 |
| Degenerated (1.5 mm) | 43.89 | 50.59 | 42.76 | 49.29 |
| Degenerated (5.0 mm) | 46.49 | 52.05 | 44.17 | 49.88 |

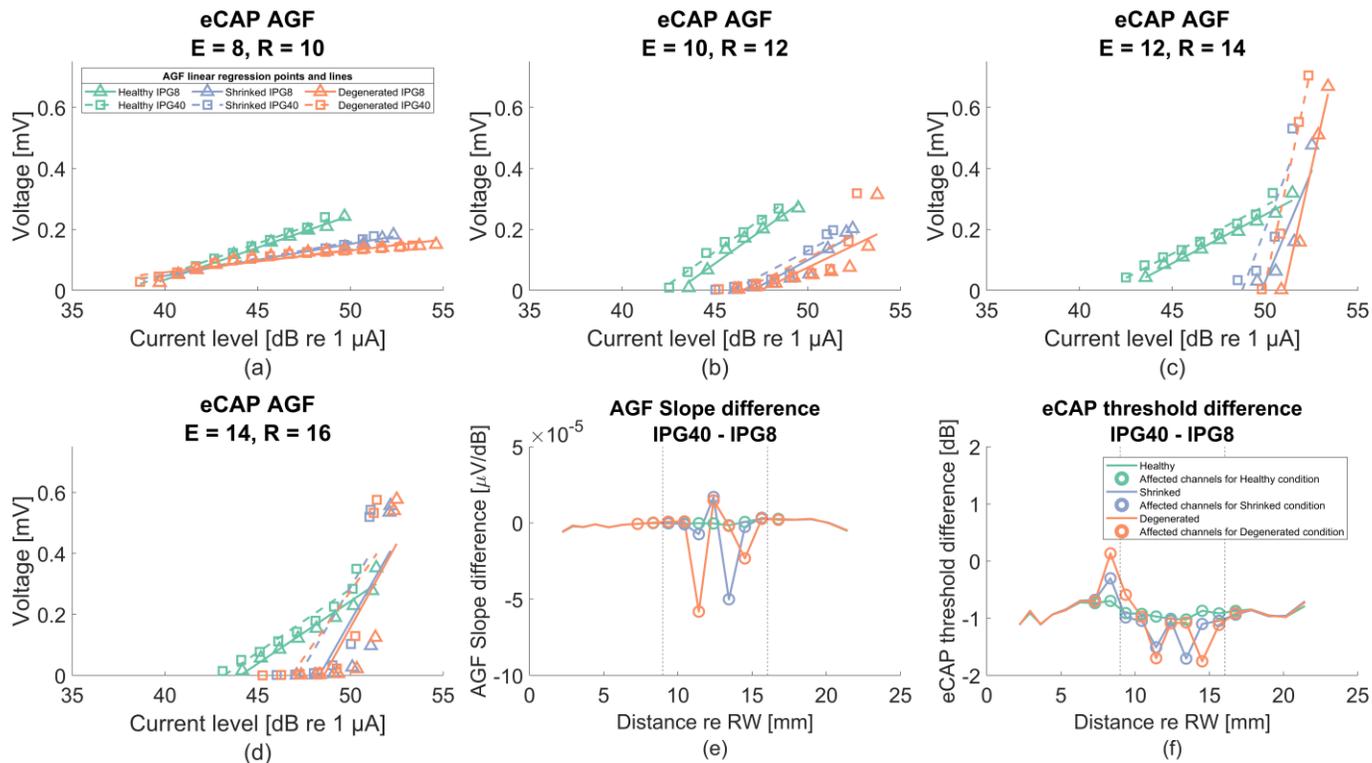

Fig.6. Analysis of the inter-phase gap (IPG) slope and IPG offset effects for electrically evoked compound action potential amplitude growth functions (eCAP AGFs), with the neural health impairment was applied to Region 6 over an impaired length of 5.0 mm (113.46°-249.16 ° re. round window). Affected channels for this special poor neural health were $E \in [7,16]$. (a-d) Predicted eCAP AGFs with linear regression lines ranging from T-level to M-level for the different neural health conditions and IPGs. Triangles represent eCAP response data points for stimuli with an 8 μs IPG (IPG8), while squares represent data points for stimuli with a 40 μs IPG (IPG40). Panels (a-d) show predicted eCAP AGFs with stimulating electrodes E8 (a), E10 (b), E12 (c), and E14(d). (e) AGF slope differences between single-pulse stimuli with IPG-8 and IPG-40 for present poor neural health condition. (f) AGF eCAP threshold differences between single-pulse stimuli with IPG-8 and IPG-40 for present poor neural health condition. Circles represent the affected channels. Different colors denote the neural health conditions, whereas solid lines represent IPG-8 and dotted lines represent IPG-40.

The first experiment compared the T-levels resulting from QP and MP stimulation modes under distinct neural health conditions. It was hypothesized that the QP stimulation would be more sensitive to poor neural health conditions. Additionally, QP stimulation was expected to show more channels with distinct T-level differences between the Healthy and impaired conditions than MP stimulation. Consistent with the hypothesis, the predicted behavioral T-levels with focused QP stimulation increased significantly proximal to the impaired region. In contrast, smaller T-level increases were observed with MP stimulation (see Fig. 3 and Fig.4(a-c)). Consistent with [9], [44], our result demonstrated that QP stimulation would generate a scaling factor with a narrower excitation peak along ANFs and provide better resolution across more channels than MP stimulation.

This study used a common criterion of exciting 1 mm of ANFs along the OC to estimate the behavioral T-level. When a channel is close to the impaired region, a higher current is required to excite the nearby ANFs with a relatively lower $I_{\text{thr},j}^{E}$ to achieve an excitation range of 1 mm (see Equation (7)). Due to the narrower excitation profile of QP stimulation, QP channels proximal to the impaired region require a higher current than MP channels to excite nearby ANFs to meet the criteria. Meanwhile, various α values with focused QP stimulation lead to scaling factors with distinct side lobes. The presence of impaired neural health conditions also excited ANFs with different side lobes at relatively lower current levels without directly stimulating impaired ANFs. This could explain the minor T-level difference between Shrinked and Degeneration conditions with focused QP mode. A consistent ∆T of 6.02 dB was observed with both stimulation modes (MP and QP) for the Shrinked condition (see Fig.4c). Since the ANFs in the Shrinked conduction did not lose their peripheral axons, the voltage distribution recorded in the shrinked ANFs was the same as in the Healthy condition. This ~6 dB upper limit observed in the experiment aligns with the assumption of a 50% reduction in the diameter of the ANFs in the Shrinked condition. According to Equation (2), $\text{AF}_{i,j}^{E}$ in Shrinked condition shows a reduction of 50% due to the decreased node diameter. This reduction leads to a 50% increase of $I_{\text{thr},j}^{E}$ base on Equation (7). The simulation results generally suggested that the behavioral T-level of focused QP stimulation was more sensitive to neural health conditions than MP stimulation.

A significant difference was found across QP and MP channels within the same neural health condition (Fig.4(d-f)). In experiments simulating poor nerve health conditions, it was found that using both QP and MP stimulation modes resulted in synchronously rising T-levels for channels proximal to the impaired region. It was hypothesized that the impaired region could be predicted based on a significant T-level difference. Our results demonstrated a significant $\Delta T_{mode}$ difference between Healthy and impaired conditions across all 12 locally impaired regions. However, for certain poor nerve health conditions such as small ILs or in Shrinked condition, channels indicated by $\Delta T_{mode}$ may not be located within the impaired region (see Suppl. Fig.4). This occurs because the relatively more minor increase in T-levels for both modes could not overcome the overall T-levels fluctuation across channels. In contrast, in poor neural health conditions within the Degenerated condition and a large IL, the channels proximal to impaired regions showed significantly higher $\Delta T_{mode}$ compared to the Healthy condition. This result suggests that a higher between-modes T-level difference could be a potential tool for neural health prediction.

The computational model was also used to predict the AGF of eCAP with two different IPGs in the range of the stimulating electrode electrical dynamic range (EDR). The model builds upon the fact that SFCC of individual ANF will vary based on the location of the stimulating electrode [34]. In the presence of impaired ANFs, the morphology of SFCC waveforms exhibited notable variations across distinct neural health conditions and IPGs, mainly when recording electrodes were positioned in proximity to the impaired region (see Fig.5e). Simulated SFCC results indicate that each ANF individually contributed to the overall CAP under varying IPG and neural health conditions, showcasing different magnitudes and peak intensities. Simulated T-levels, M-levels, and resulting EDR, shown in Table I, varied in different neural health conditions and IPGs. Furthermore, the T-levels and M-levels in Healthy condition with an IPG of 8 µs were significantly higher than those with an IPG of 40 µs. Increase and IPG between phases of biphasic stimulus would have more probability to evoke spikes in the first phase before being balanced by the second phase, which has been reported in previous studies [5], [16], [17], [45]. Under poor neural health conditions, channels required more current to excite ANFs, consistent with the findings of the first experiment.

The eCAP waveform predicted from the aforementioned results closely resembles the clinically recorded human eCAP waveform. Similar to [34], a secondary negative peak followed by the N1 peak was observed, regardless of the specific neural health condition (see Fig. 5g). A more realistic human ANF morphology, particularly the peripheral processes and soma, could achieve more accurate predictions of SFCCs and eCAPs.

For experiment 2, it was hypothesized that IPG slope and IPG offset effects would reduce when ANFs were in poor neural health conditions. Unlike previous studies [5], simulation results indicated that AGF $\Delta$Slope did not follow a consistent pattern in each recording electrode, regardless of the neural health condition. These inconsistent conclusions may be attributed to the different units of input current and the fitting process. An additional linear fitting test with linear input current units (i.e., µA) was applied, and the observed behavior of the AGF slope aligns with animal experimental findings, showing an increase in AGF slope as the IPG increases (not shown). Similar to [23] and [46], simulation eCAP results failed to find a uniform $\Delta$Slope pattern for affected channels in poor neural health conditions. Our model may not have been sufficiently sensitive to detect an IPG slope effect reduction in impaired neural health conditions implemented with IL of 0.5 mm and 1.5 mm (see Suppl. Fig.4(a-b)). Furthermore, a notable decrease toward negative infinity in AGF $\Delta$Slope was observed in the Degenerated condition with an IL of 5 mm compared to the Healthy condition. In this scenario, both IPGs exhibited higher EDR, crossing the lower asymptote of the AGF and entering the linear increase region, as shown in Fig.6c. One possible explanation is the unrealistic estimation of the M-level. The 4 mm excited OC length criterion for M-level prediction was a simple solution suggested by previous studies [37], [47], [48], but it may overestimate the M-level. Another possible reason is that the linear regression of AGF does not have a plausible estimation of AGF slope, as shown in Fig.6d. The unrealistic estimation of AGF linear slope may also contribute to the discrepancy pattern of across-channel $\Delta$Slope.

The simulation results also demonstrate a distinct decrease in AGF eCAP threshold with increasing IPG, irrespective of neural health conditions. This finding is consistent with reported observations in animal experiments and clinical reports [5], [15], [21]. This behavior might reflect the intrinsic properties of ANFs, where varying current intensities are required to elicit stimulation at different IPGs. The predicted eCAP threshold closely matched the T-level of the channel, and a significant AGF $\Delta$Threshold decrease was observed across neural health conditions in affected channels with ILs of 1.5 and 5.0 mm. Consistent with the experiment 1, a significant difference in AGF $\Delta$Threshold between the Shrinked and Degenerated conditions was observed only when the IL was 5 mm. This occurred because the T level in the Shrinked condition reached the upper limit of ~6 dB. As suggested by [49], it may be possible to assist in predicting the T-level of a channel by linear fitting the AGF. However, eCAP threshold differences across channels were highly variable and did not reveal a consistent pattern across neural health conditions (see Suppl. Fig. 5). The linear regression implemented in the study was applied to data points with recorded non-zero eCAP amplitude. In some channels, corresponding T-levels were sufficient to stimulate the ANF and produce a smaller eCAP amplitude (see Fig.6d). Including these data points in the fitting process may lead to variations in the $\Delta$Threshold across channels. Including these data points in the fitting may have caused variations in $\Delta$Threshold across channels. Moreover, when stimulating electrodes near impaired regions (e.g., E=8, with impaired ANFs in Region 6 and IL=5.0 mm, see Fig. 6a), the T-levels did not vary significantly across neural health conditions due to the broader stimulation range of MP mode and fewer impaired ANFs near the electrode. Consequently, the previously described AGF morphology exhibited observable changes only in the maximum amplitude. According to Equation (7), long IPG stimulation results in lower $I_{thr}$ and T-levels than short IPG stimulation, due to an increased opportunity to stimulate ANFs. In this scenario, the T-levels predicted by long IPG stimulation were close to those in the Healthy conduction. In contrast, short IPG stimulation was more sensitive to poor neural health conditions, increasing T-levels. This T-level discrepancy could explain the abnormal phenomenon where the channel $\Delta$Threshold was greater than zero at the boundary of the impaired region.

Comparing eCAP AGF slopes with previous studies posed challenges due to differences in species, unknown neural health conditions, variations in data recording techniques, and disparate data postprocessing fitting methods. Several factors may have contributed to the variability observed in the eCAP threshold and AGF slope results. First, the present experiment 2 focuses on the demyelination and degeneration of peripheral processes. Despite impairment in each ANF, there has been no loss of soma, meaning the SGN density has remained unchanged over time. Animal studies by Prado-Guitierrez et al. (2006) and Ramekers et al. (2014) suggested that there is a correlation between surviving SGN density and the IPG offset effect [5], [16]. Moreover, Brochier et al. (2021) further demonstrated this correlation through a simple theoretical model and a reanalysis of animal and human eCAP data that measured distinct IPGs [30]. A recent human study examining the impact of IPG offset on children with cochlear nerve deficiency (CND) found that CND children exhibited less of a decrease in eCAP threshold with increased IPG compared to those with normal-sized ANFs [50], [51]. The findings of Schvartz-Leyzac and Pfingst (2016) showed that the AGF linear slope increases with rising IPG when using a linear input current unit for adult CI subjects [46]. However, Hughes et al. (2018) reported that no IPG slope effect was observed in adult CI subjects when linear regression was implemented using an input unit of dB [52]. Similarly, Jahn and Arenberg measured AGF linear slope difference between aged CI subjects using a dB input current unit. Their data suggested that children CI subjects (i.e., better neural health CI users) had a steeper AGF liner slope. However, no significant change in AGF slope across subjects was observed with increasing IPG [21], [22]. Meanwhile, the influence of the stimulating and recording electrodes' positions relative to the dead region remains unknown. We observed some discrepancy patterns of the AGF slope and threshold difference with the stimulating electrodes at the impaired region border. However, large and inconsistent data fluctuations across different impaired regions have hindered our ability to detect this influence. Due to the inconsistent results across different studies, the reliability of the IPG slope and offset effects remains uncertain.

## V. Conclusion

This study utilized a 3D computational model of the stimulated human cochlea to evaluate two approaches for predicting neural health: focused stimulation and IPG slope/offset effects on eCAP AGFs. This model effectively simulated ANF excitation profiles under various neural health conditions. Simulation results indicate that the IPG slope and IPG effects with a linear-logarithmic linear fitting are not a robust estimation method of neural health. In contrast, larger changes in T-levels were observed for MP and QP stimulation in electrodes adjacent to the impaired region. These findings suggest that higher between-mode T-level differences could serve as a potential predictive tool for assessing impaired neural health. These measures show promise for predicting neural health. Additionally, focused QP stimulation exhibited a relatively higher shift in predicted behavioral thresholds with a small impaired length of 0.5 mm, highlighting its potential advantage for neural health estimation in regions with minor impairments. This computational model enhances our understanding of neural excitation profile differences across various neural health conditions. Future research could explore non-neural factors such as CI position, and cochlear geometry characteristics could also provide valuable insights for developing more accurate neural health estimation methods.

## VI. Data Available

The raw data of simulated T-levels, M-levels of MP and QP stimulations, and the eCAP are available in Gitlab: https://gitlab.gwdg.de/apg/neural-health.

# Supplementary Material for:

# Evaluating electrophysiological and behavioral measures of neural health in cochlear implant users: a computational simulation study

Yixuan Zhang, Daniel Kipping, and Waldo Nogueira

## VII. Overview

This supplementary material contains details on the stimulation modes and additional results for the computational experiments.

Suppl.Tab.1. provides a detailed collection of the cochlear implant (CI) electrode array diameters at each electrode contact center, the electrode insertion angles, and the distance along the organ of Corti in relation to the round window (RW) of the CI 632 geometry used in this study.

Suppl.Tab.2. provides a detailed description of the neural health conditions used in our study. This includes a healthy condition ("Healthy") and two impaired conditions ("Shrinked" and "Degenerated"). Each impaired condition was applied over a length of 0.5 mm, 1.5 mm, and 5.0 mm in one out of 12 local impaired regions, linearly spaced at 30-degree intervals along the cochlea.

**Experiment 1: Quadrupolar (QP) vs. Monopolar (MP) threshold**
Supp.Fig 1 displays the simulated behavioral thresholds (T-levels) for MP and QP stimulations across the insertion distance of the channels within the 12 impaired regions and an impaired length of 0.5 mm (a), 1.5 mm (b), and 5.0 mm (c).

Supp.Fig.2 shows the distribution of residual T-level differences between QP and MP stimulation across channels for each neural health condition within an impaired length of 0.5 mm (a), 1.5 mm (b), and 5.0 mm (c).

**Experiment 2: The eCAP slope and offset effects**
Suppl. Fig. 3 illustrates one example of the predicted amplitude growth function (AGF) for different neural health conditions and IPGs. The stimulation current range to estimate the eCAP AGF started from 5 dB below T-level to M-level in steps of 1 dB for each stimulating electrode $E$. The recording electrode $R$ was defined as the stimulating electrode $E$ plus two (+2) in the apical direction. The shown example assumes an impaired region occurred at Region 6 (positioned at approximately 180° relative to the RW) within an impaired length (IL) of 0.5 mm (a), 1.5 mm (b), and 5 mm (c).

Suppl. Fig. 4 and 5 present AGF slope and eCAP threshold difference between two single pulse stimuli with IPGs of 8 μs (IPG-8) or 40 μs (IPG-40) for experiment 2, respectively. The shown data assumes an impaired region occurred within an impaired length (IL) of 0.5 mm (a), 1.5 mm (b), and 5 mm (c). The x-axis represents the distance of stimulating electrode $E$ relative to the RW.

| ID | CI Array Diameter [mm] | Angle re RW [°] | Distance re RW [mm] |
|---|---|---|---|
| Ebase | 0.475 | 0 | 0 |
| E1 | 0.475 | 25.43 | 2.21 |
| E2 | 0.469 | 33.53 | 2.91 |
| E3 | 0.4631 | 41.73 | 3.62 |
| E4 | 0.4571 | 51.19 | 4.43 |
| E5 | 0.4512 | 62.24 | 5.33 |
| E6 | 0.4452 | 74.62 | 6.30 |
| E7 | 0.4393 | 88.09 | 7.29 |
| E8 | 0.4333 | 103.45 | 8.33 |
| E9 | 0.4274 | 119.59 | 9.36 |
| E10 | 0.4214 | 137.86 | 10.44 |
| E11 | 0.4155 | 155.73 | 11.41 |
| E12 | 0.4095 | 175.61 | 12.41 |
| E13 | 0.4036 | 197.43 | 13.44 |
| E14 | 0.3976 | 219.96 | 14.52 |
| E15 | 0.3917 | 242.05 | 15.65 |
| E16 | 0.3857 | 262.74 | 16.77 |
| E17 | 0.3798 | 281.73 | 17.82 |
| E18 | 0.3738 | 303.61 | 18.99 |
| E19 | 0.3679 | 326.95 | 20.16 |
| E20 | 0.3619 | 354.92 | 21.44 |
| E21 | 0.3560 | 381.79 | 22.64 |
| E22 | 0.35 | 404.55 | 23.73 |

Suppl. Tab. 1 Inter-electrode array diameters, the electrode contact insertion angle, and distance along the organ of Corti re. RW for used CI632 geometry. The CI array was modeled as a simplified spiral cylinder with a linearly decreasing radius, starting with a diameter of 0.475 mm at the basal end and tapering to 0.3 mm at the apical end.

| Condition Type | ANF state | Impaired length [mm] | Local Dead Regions |
|---|---|---|---|
| Healthy | Healthy | N/A | N/A |
| Impaired | Shrinked | 0.5 | 12 (Linear spacing at 30° intervals) |
| Impaired | Shrinked | 1.5 | 12 (Linear spacing at 30° intervals) |
| Impaired | Shrinked | 5 | 12 (Linear spacing at 30° intervals) |
| Impaired | Degenerated | 0.5 | 12 (Linear spacing at 30° intervals) |
| Impaired | Degenerated | 1.5 | 12 (Linear spacing at 30° intervals) |
| Impaired | Degenerated | 5 | 12 (Linear spacing at 30° intervals) |

Suppl. Tab. 2  Neural Health Conditions Used in the Study

IL = 0.5 mm

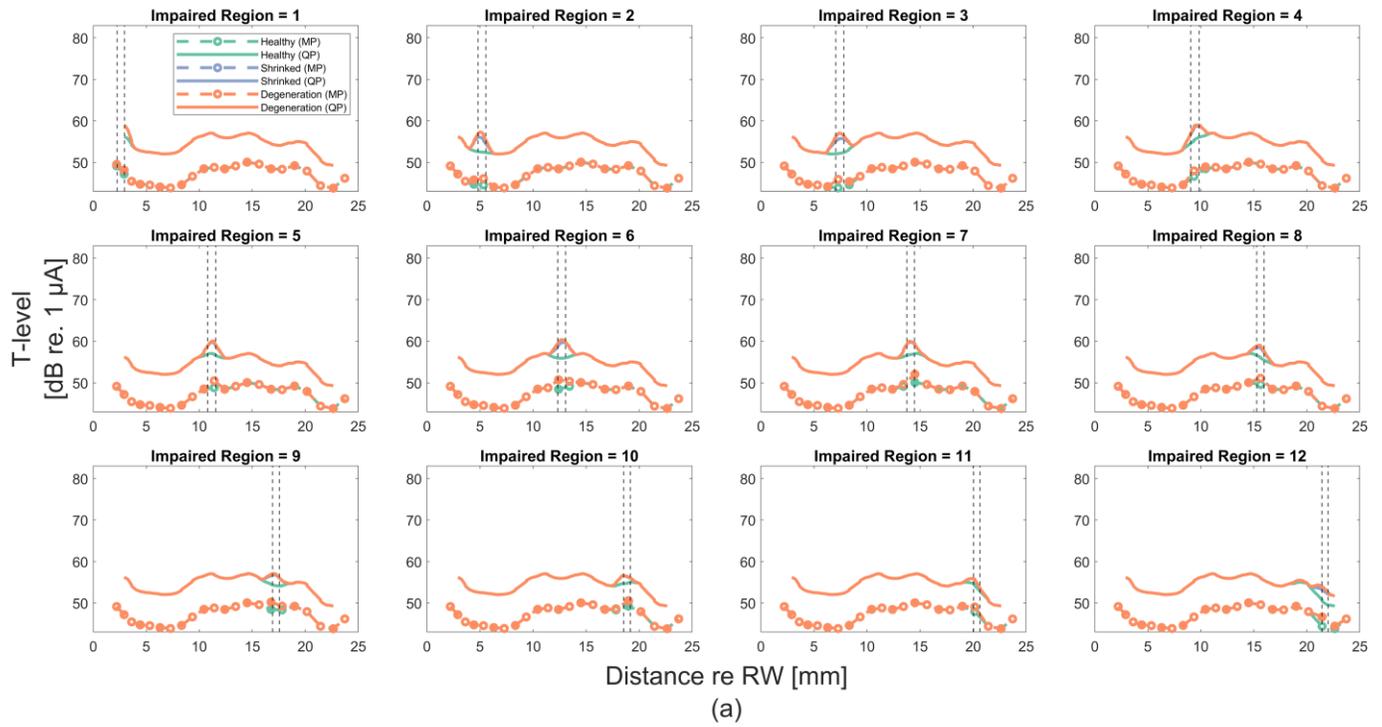

(a)

IL = 1.5 mm

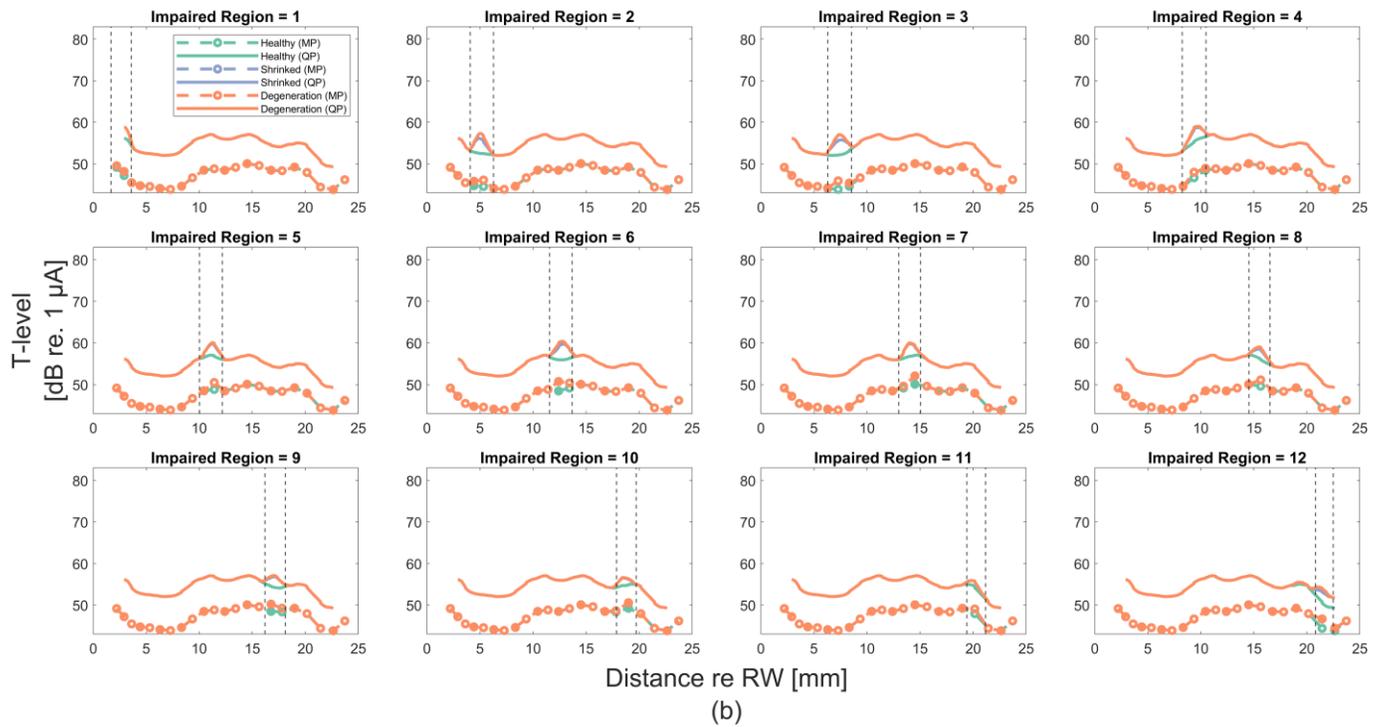

(b)

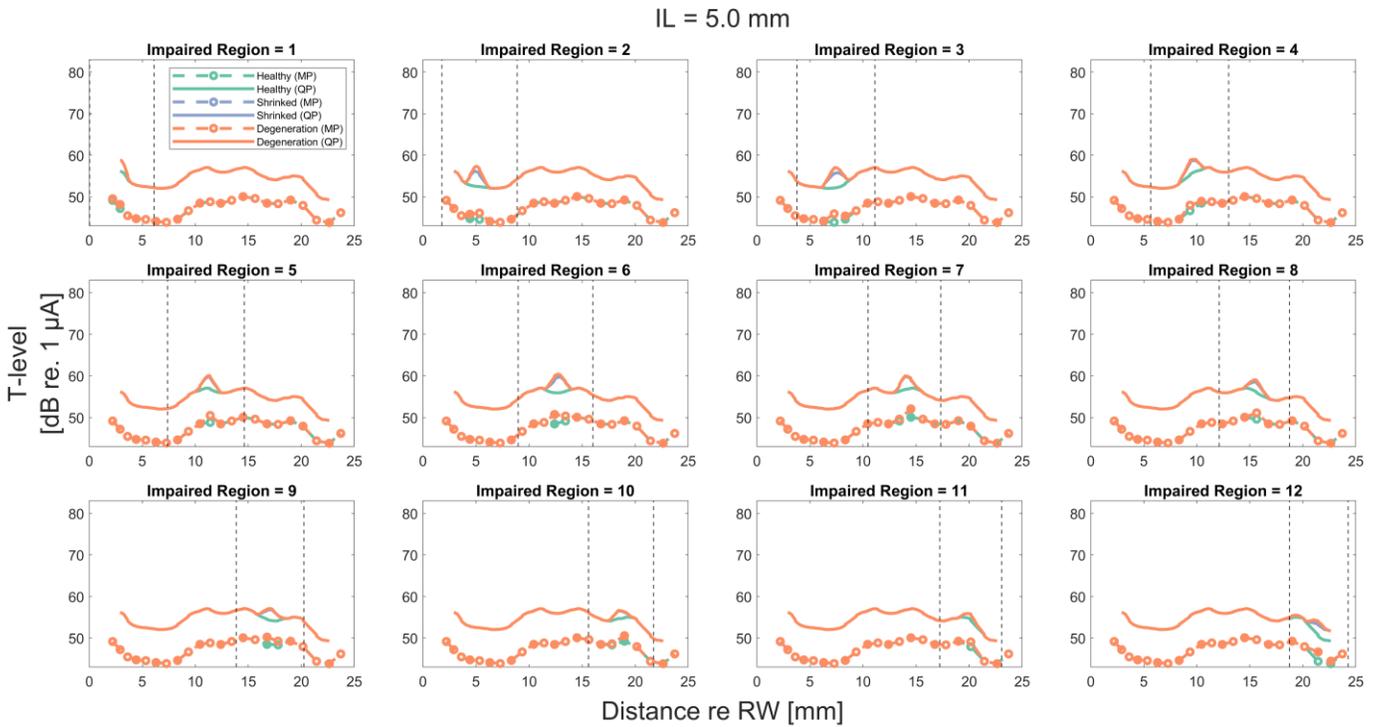

Suppl. Fig. 1. Estimated behavioral T-level profiles for monopolar (MP) and quadrupolar (QP) stimulation in different neural health conditions applied to the impaired regions 1 to 12 over an impaired length of 0.5 mm (a), 1.5 mm (b). and 5.0 mm (c). The x-axis shows the distance of insertion of each MP or QP channel relative to the round window (RW). The dotted black vertical lines indicate the borders of the impaired region. The QP stimulation results are displayed as a moving average across 11 neighboring channels.

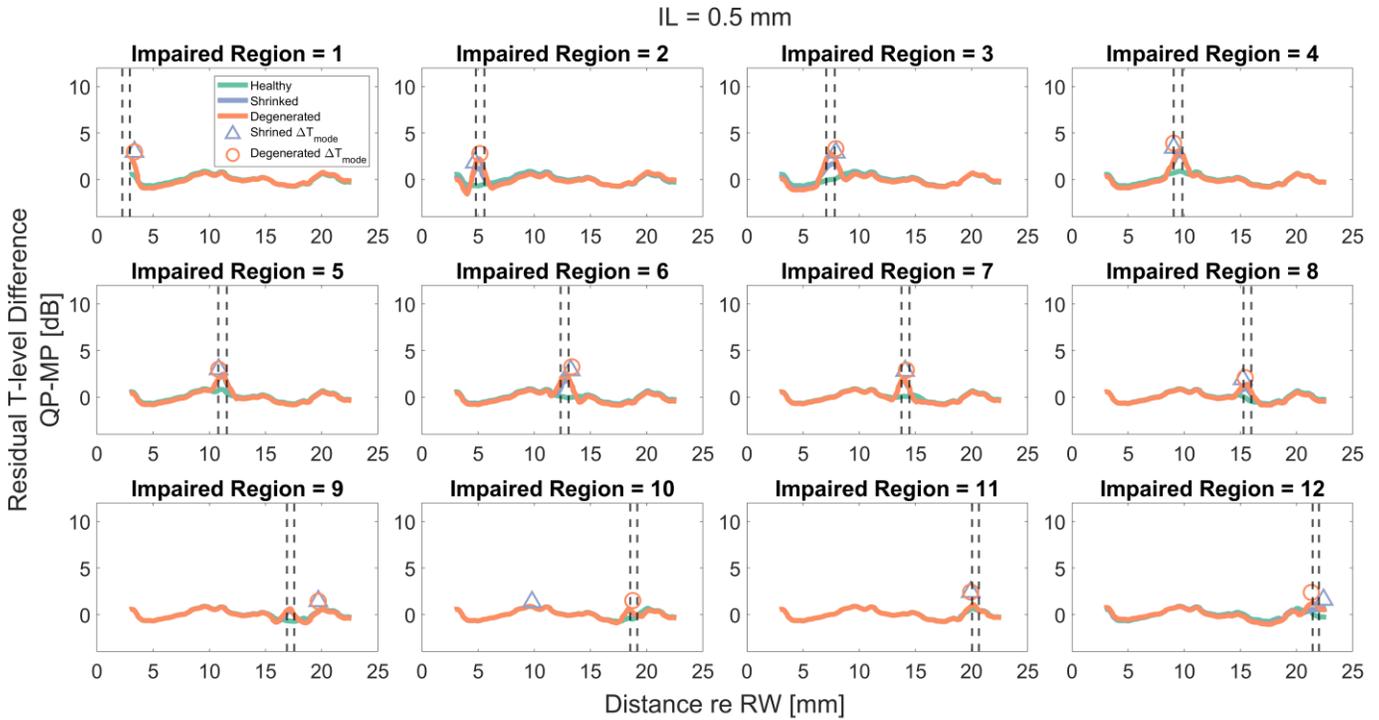

IL = 1.5 mm

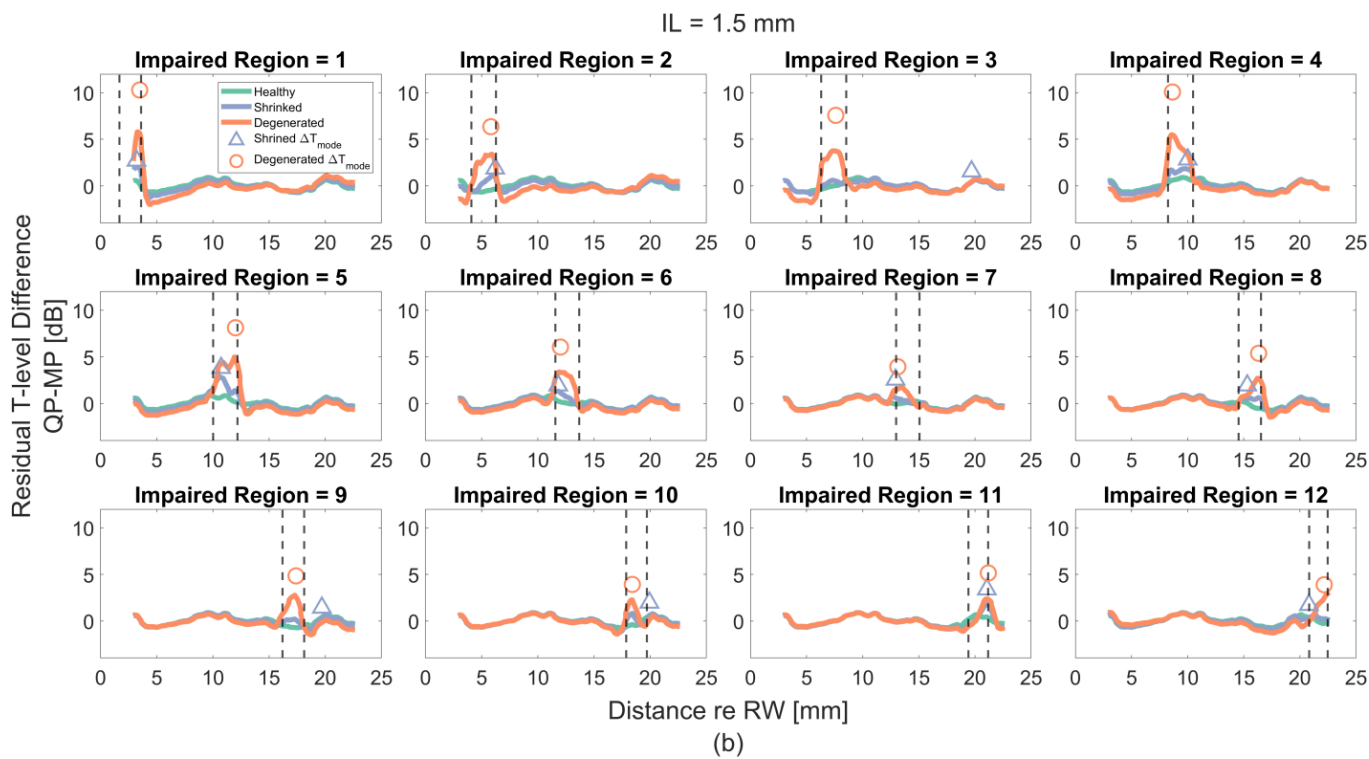

(b)

IL = 5.0 mm

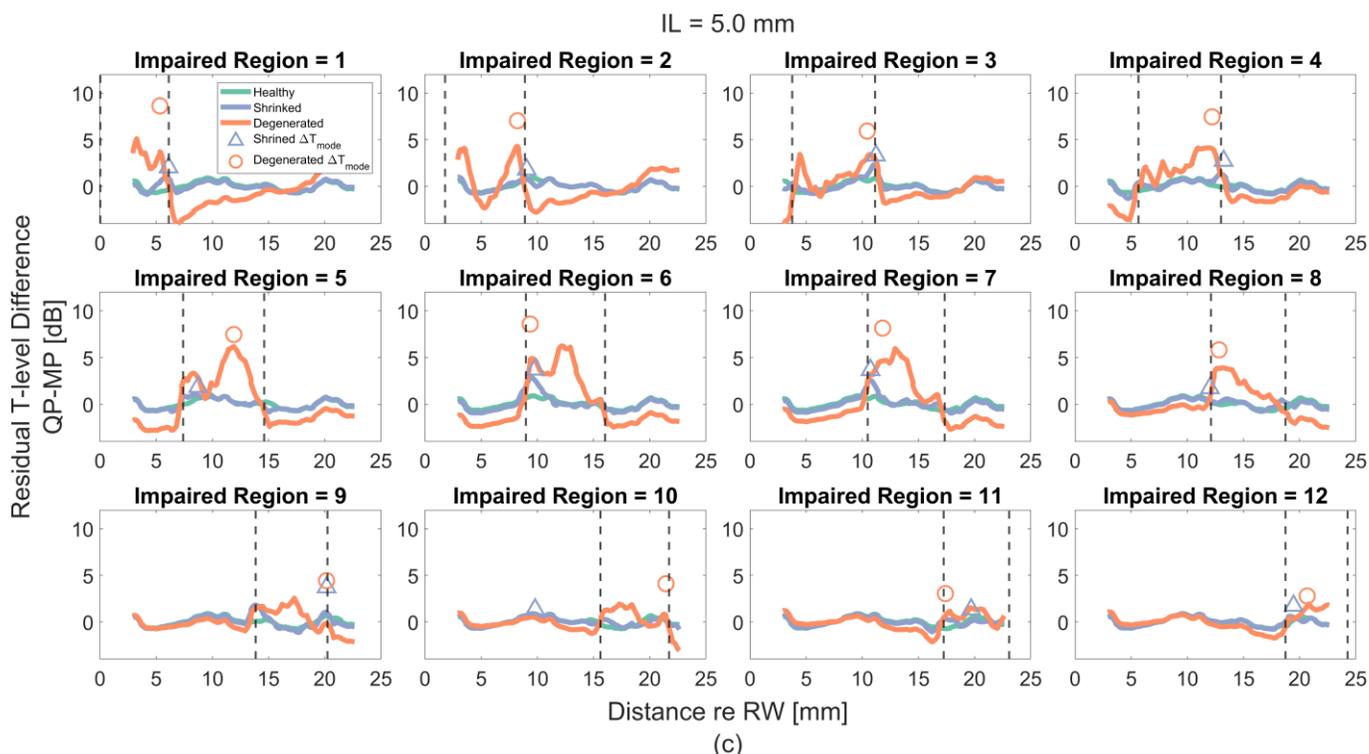

(c)

Suppl. Fig. 2 Distribution of residual T-level differences between quadrupolar (QP) and monopolar (MP) stimulation modes in different neural health conditions. The ANF impairment occurred in one of 12 local impaired regions over an impaired length of 0.5 mm (a), 1.5 mm (b), or 5.0 mm (c). The maximum residual T-level difference of each neural health condition was marked as $\Delta T_{mode}$. The dotted black vertical lines indicate the borders of the impaired region. The residual T-level differences are displayed as a moving average across 11 neighboring channels.

IL = 0.5 mm

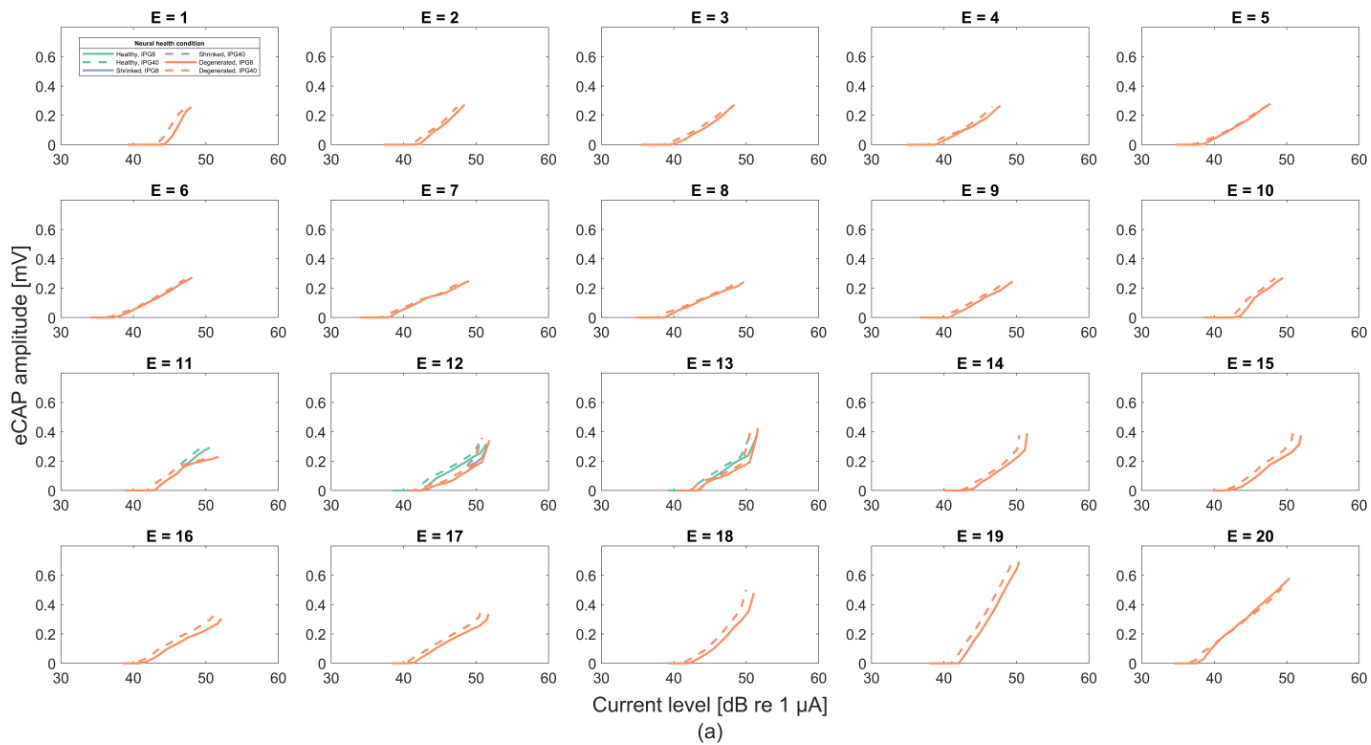

(a)

IL = 1.5 mm

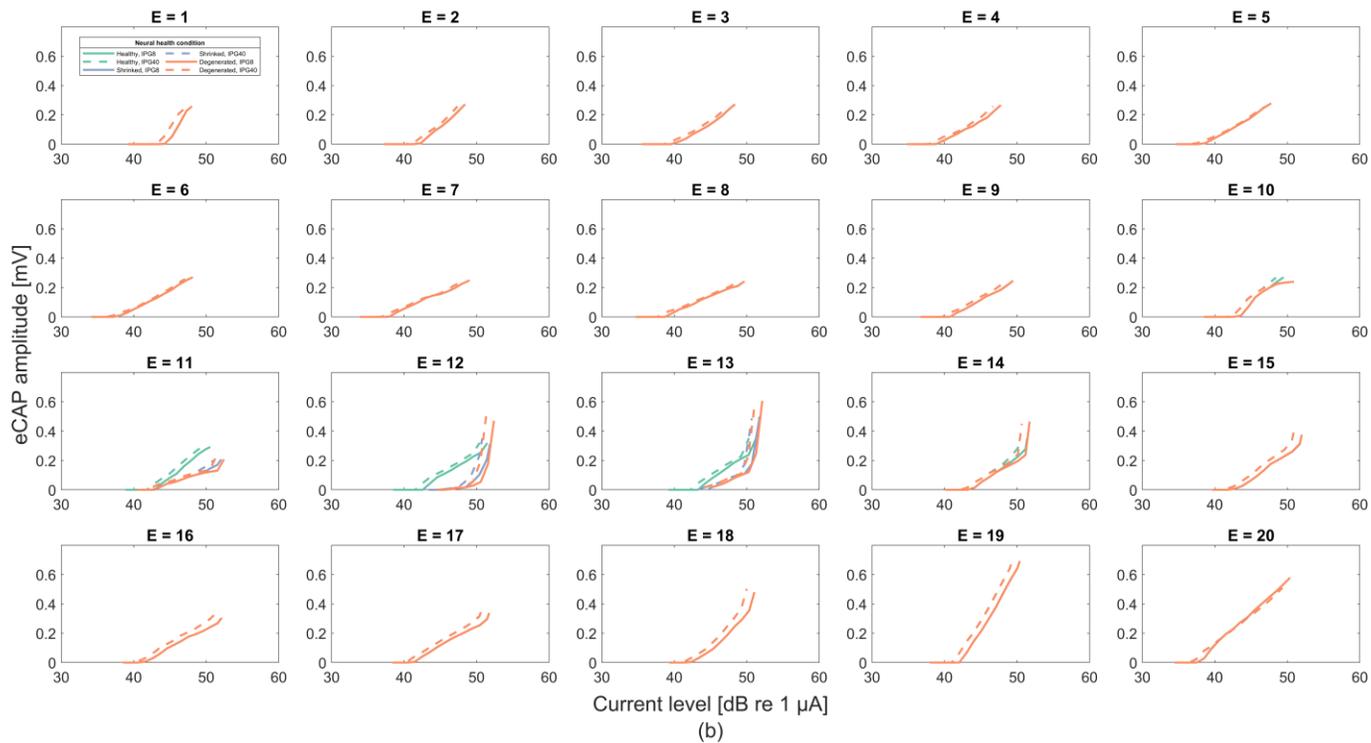

(b)

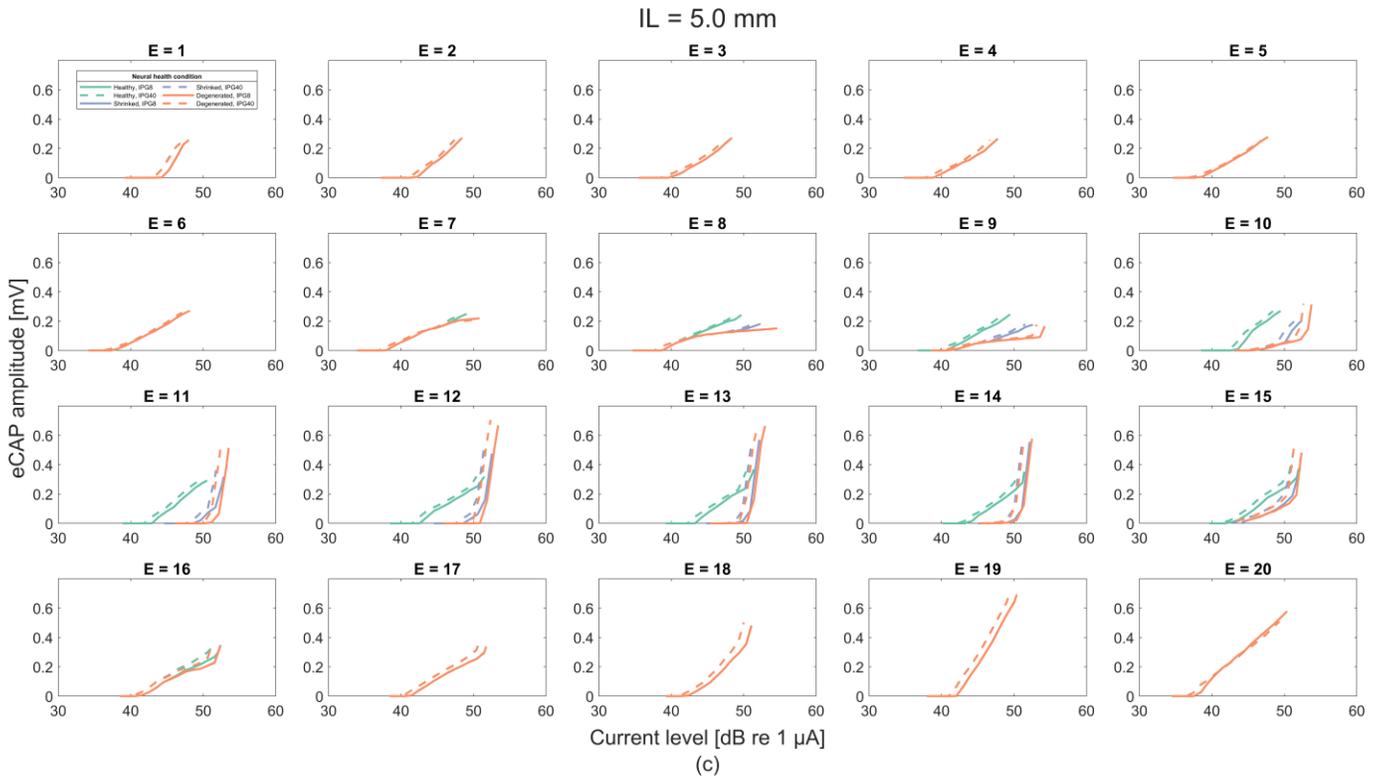

Suppl. Fig. 3 Examples of predicted electrically evoked compound action potential amplitude growth functions (eCAP AGFs) for different neural health conditions and inter-phase gaps (IPG). The stimulation current range to estimate the eCAP AGF started from 5 dB below T-level and ranged up to M-level in steps of 1 dB for each stimulating electrode $E$. The recording electrode $R$ was defined as the stimulating electrode $E$ plus two (+2) in the apical direction. The shown example results from an impaired neural health applied to region 6 (positioned at approximately 180° relative to the round window close to electrode E12) over an impaired length of 0.5 mm (a), 1.5 mm (b), or 5.0 mm (c). Different colors denote the neural health conditions, whereas solid lines represent IPG-8 and dotted lines represent IPG-40.

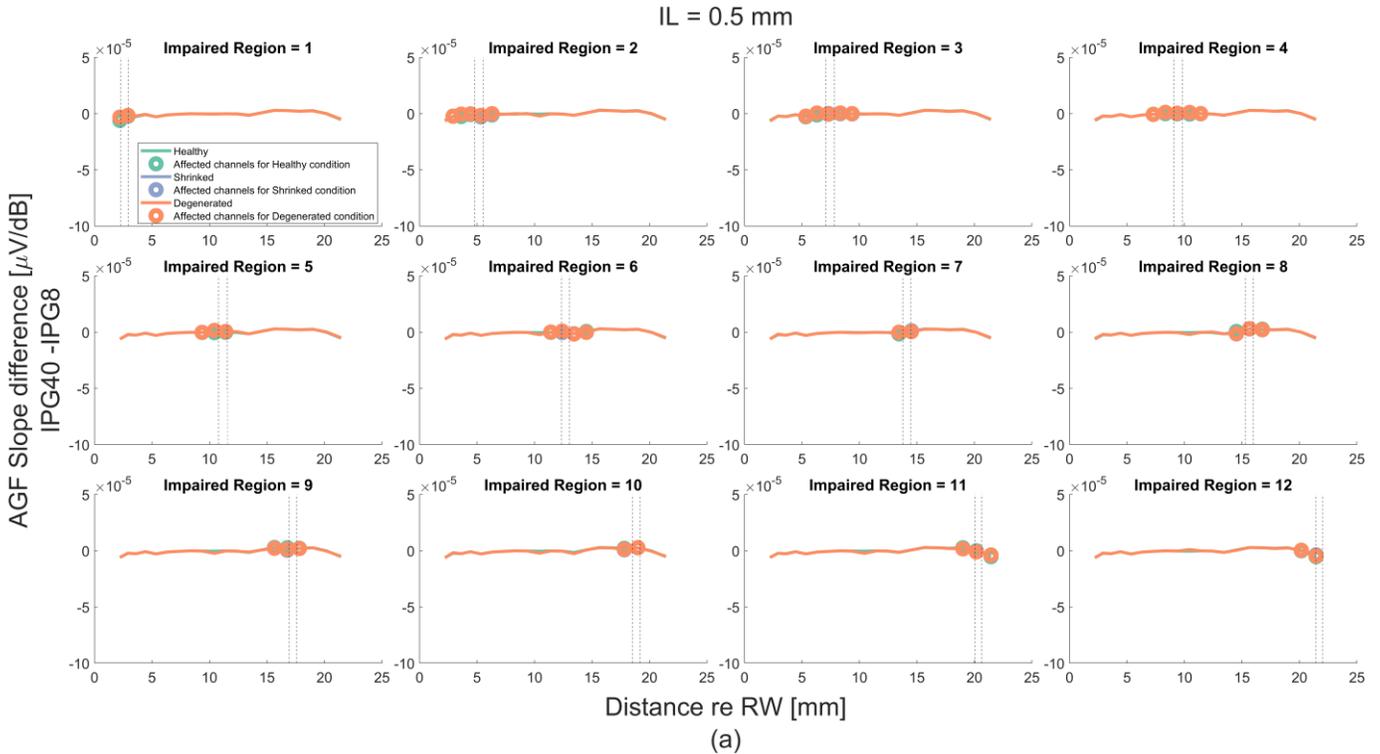

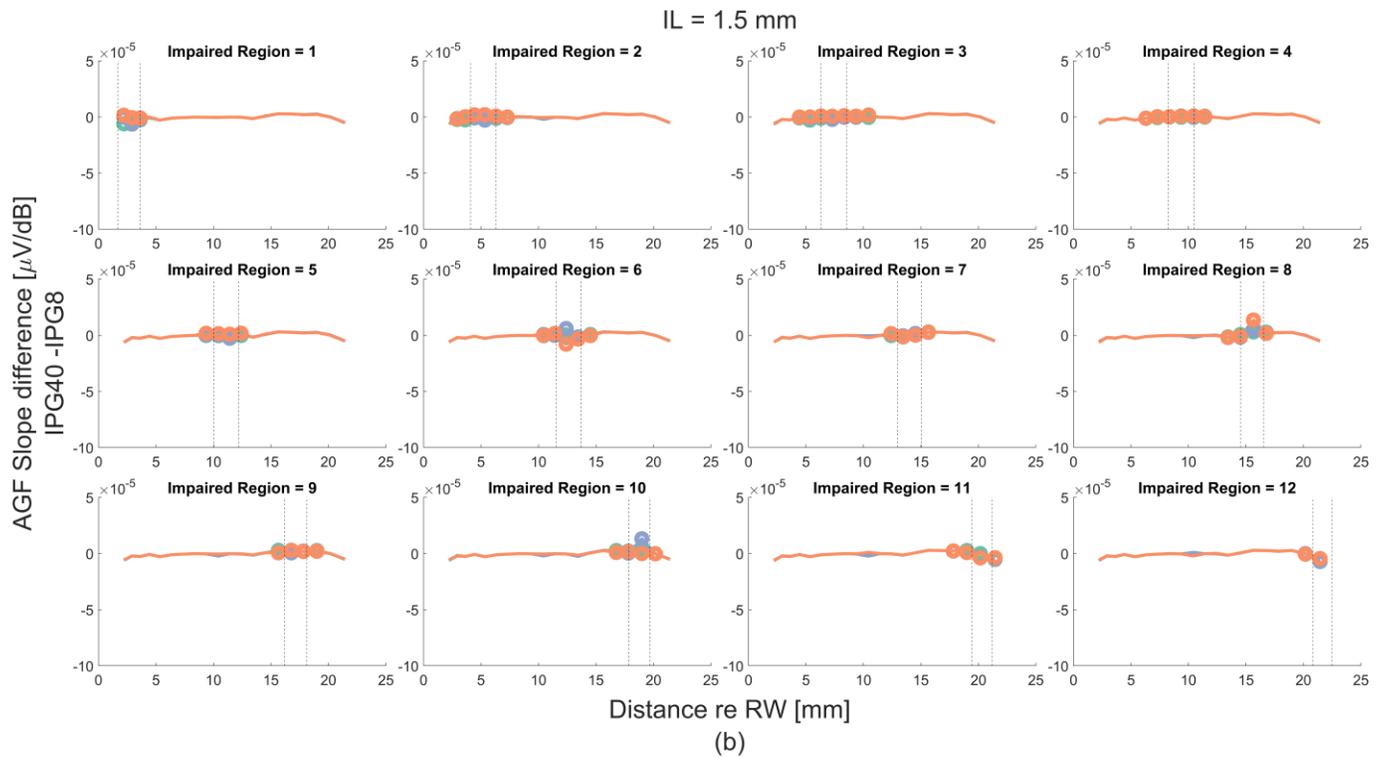

(b)

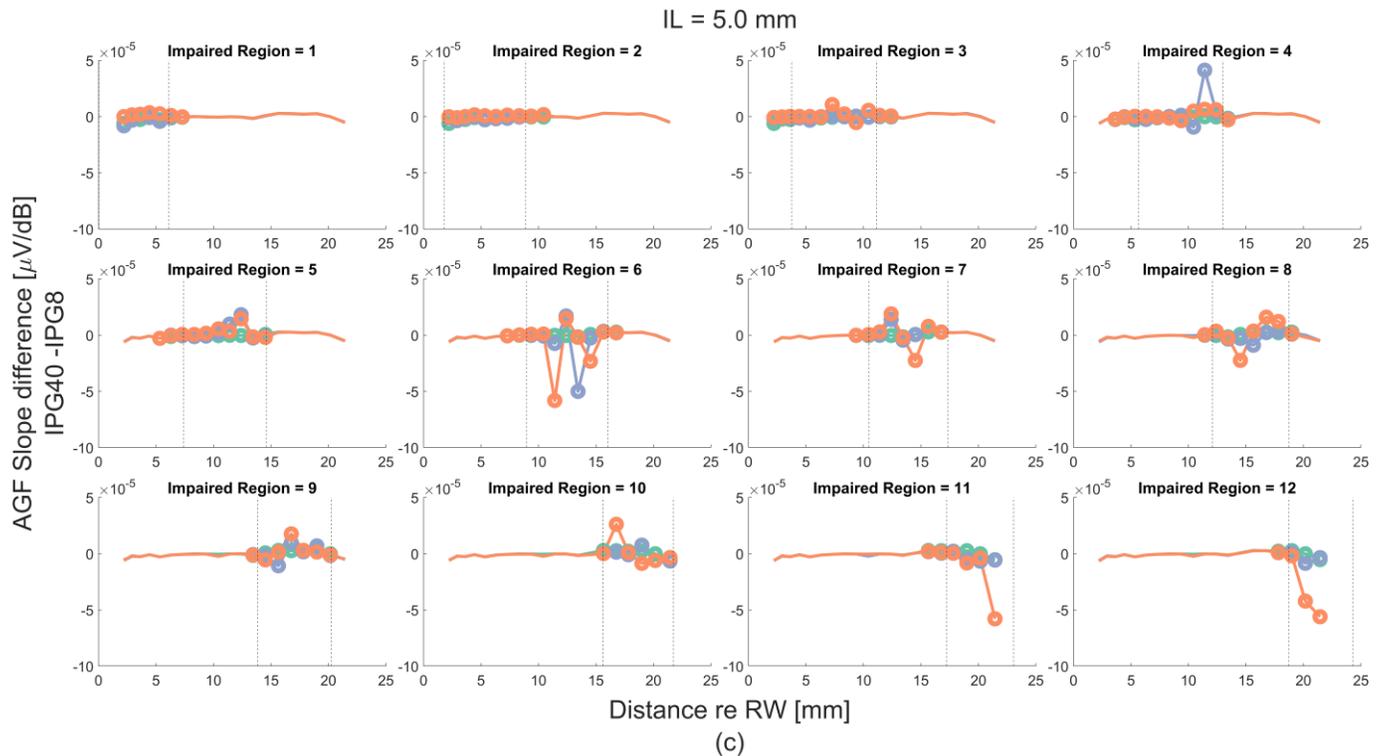

(c)

Suppl. Fig. 4 AGF slope differences between single-pulse stimuli with IPG-8 and IPG-40 for Experiment 2. The data assumes an impaired region within an impaired length (IL) of 0.5 mm (a), 1.5 mm (b), and 5 mm (c). Different colors denote the neural health conditions, whereas solid lines represent IPG-8 and dotted lines represent IPG-40. Circles represent the affected channels. The x-axis represents the stimulating electrode's distance relative to the round window (RW).

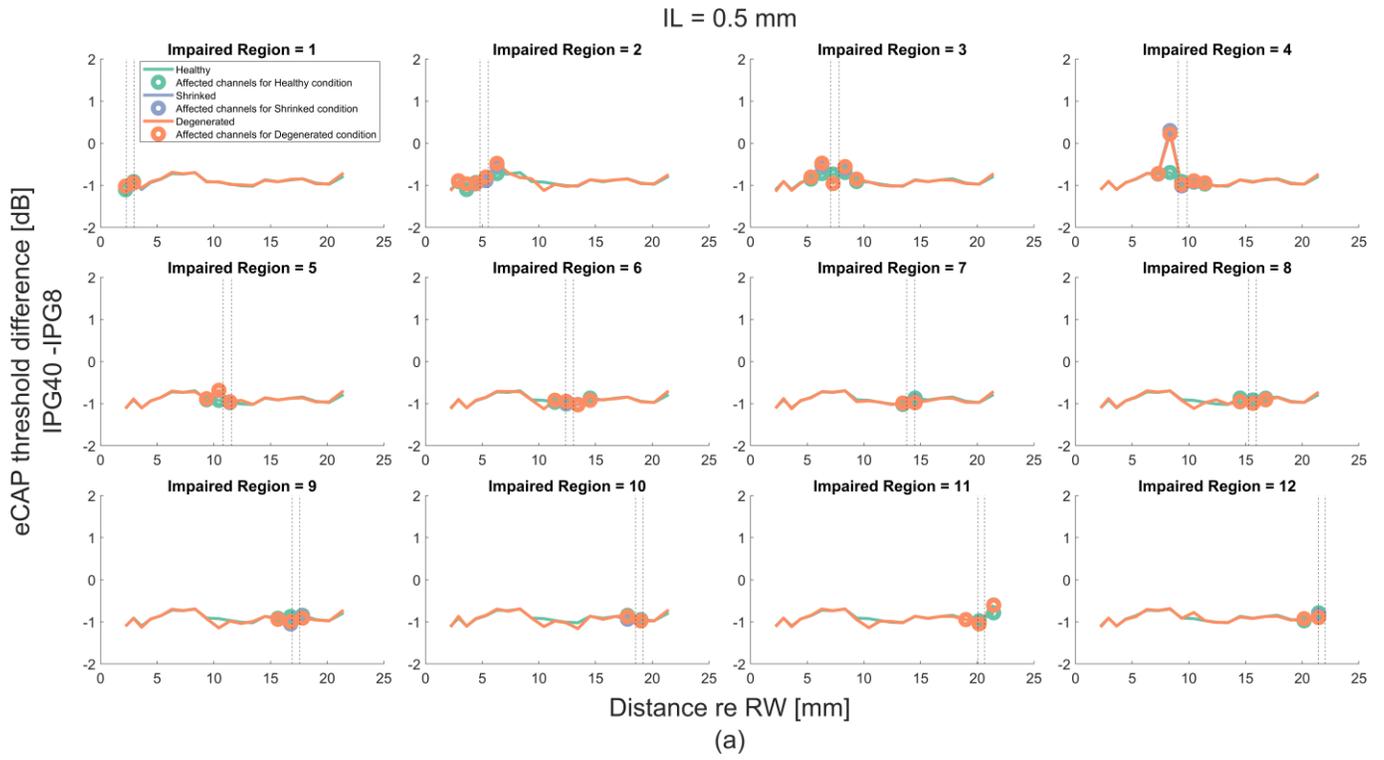

(a)

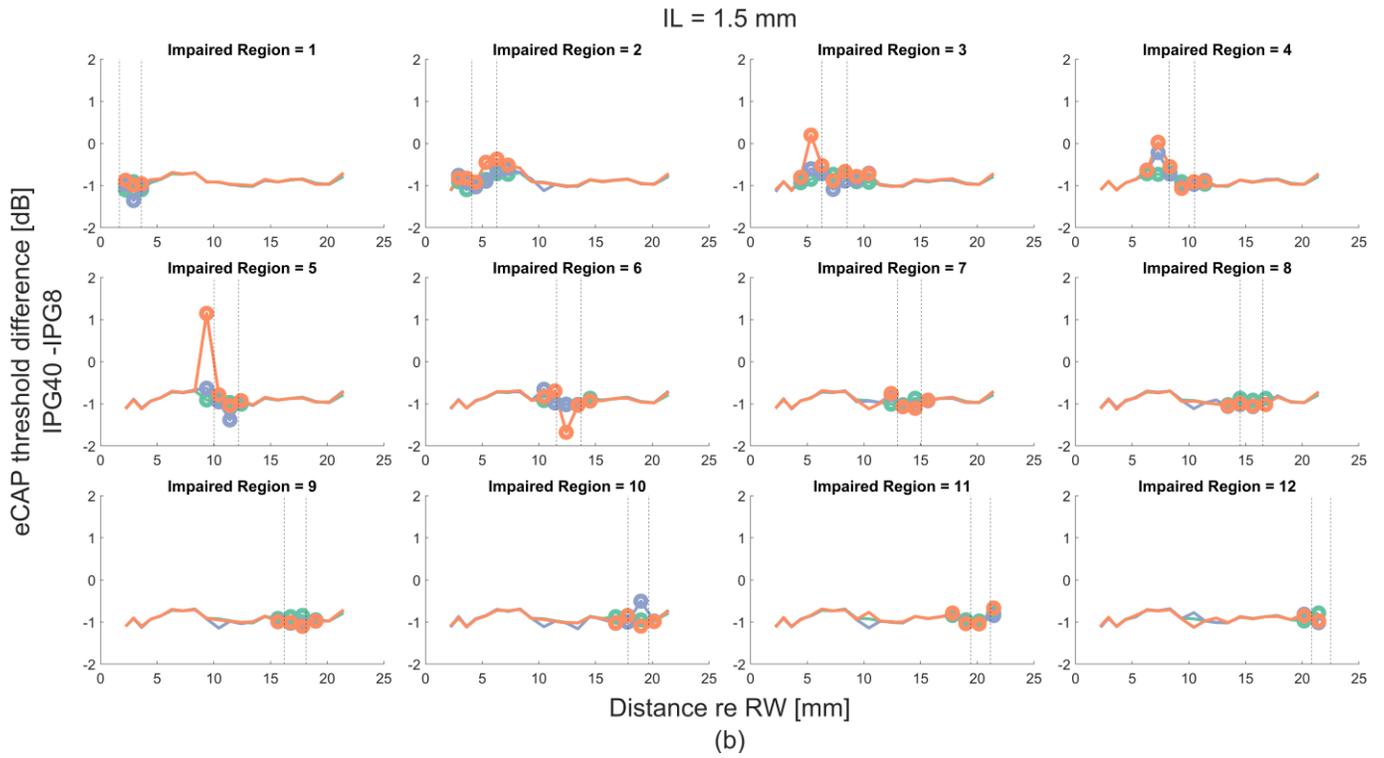

(b)

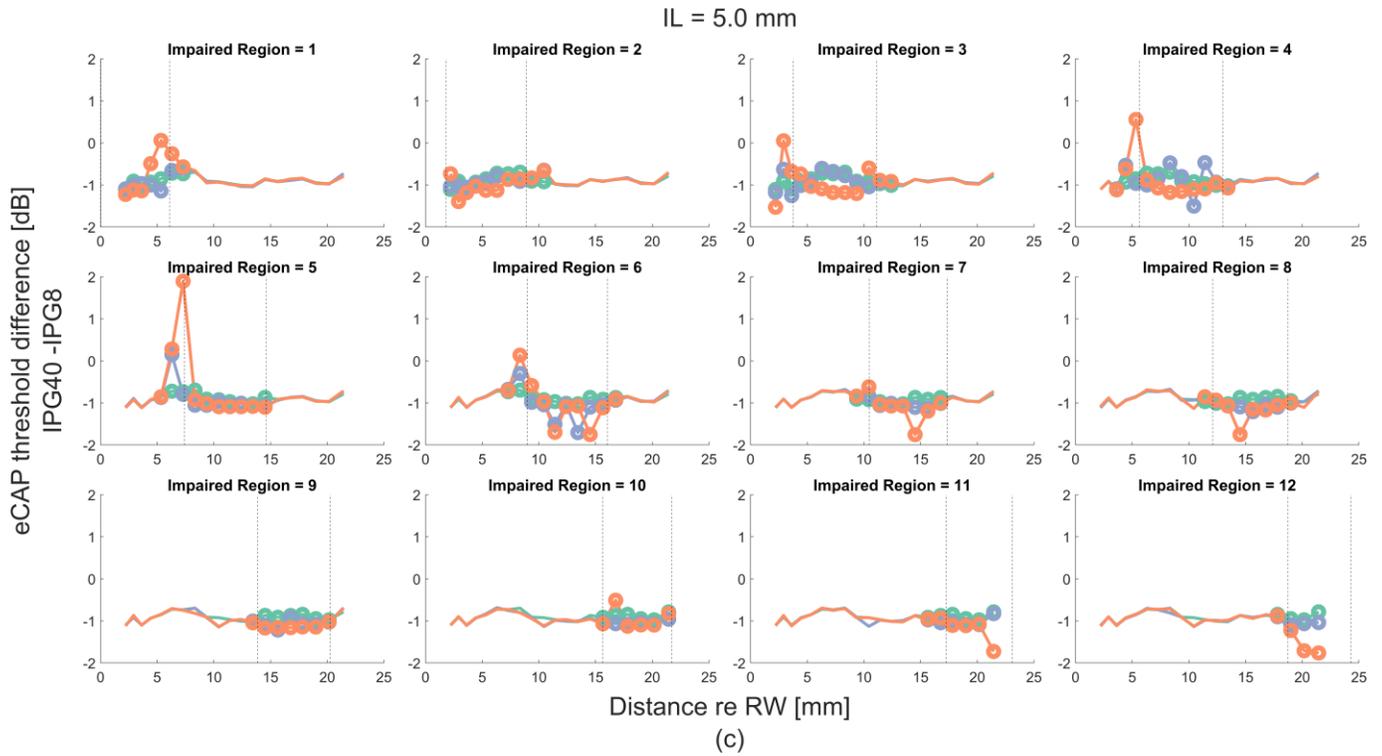

Suppl. Fig. 5 AGF threshold differences between single-pulse stimuli with IPG-8 and IPG-40 for Experiment 2. The data assumes an impaired region within an impaired length (IL) of 0.5 mm (a), 1.5 mm (b), and 5 mm (c). Different colors denote the neural health conditions, whereas solid lines represent IPG-8 and dotted lines represent IPG-40. Circles represent the affected channels. The x-axis represents the stimulating electrode's distance relative to the round window (RW).